\documentclass{aa}
\usepackage{graphicx}
\usepackage{txfonts}

\usepackage[switch]{lineno} 
\usepackage{amsmath}    
\usepackage{amssymb}    
\usepackage{cases}  
\usepackage{multirow} 
\usepackage{xcolor} 

\usepackage{enumitem}
\setlist[itemize]{topsep=0pt,after=\vspace{.5\baselineskip}}

\DeclareRobustCommand*{\drv}{\mathop{}\!\mathrm{d}}
\DeclareRobustCommand*{\kms}{km\,s$^{-1}$}

\begin{document} 

   \title{Solar wind charge exchange in cometary atmospheres}

   \subtitle{I. Charge-changing and ionization cross sections for He and H particles in H$_2$O}

   \author{Cyril Simon Wedlund\inst{1}
\and Dennis Bodewits\inst{2}
\and Markku Alho\inst{3}
\and Ronnie Hoekstra\inst{4}
\and Etienne Behar\inst{5,6}
\and Guillaume Gronoff\inst{7,8}
\and Herbert Gunell\inst{9,10}
\and Hans Nilsson\inst{5,6}
\and Esa Kallio\inst{3}
\and Arnaud Beth\inst{11}
}

   \institute{Department of Physics, University of Oslo, P.O. Box 1048 Blindern, N-0316 Oslo, Norway\\
             \email{cyril.simon.wedlund@gmail.com}
        \and 
            Physics Department, Auburn University, Auburn, AL 36849, USA
        \and
            Department of Electronics and Nanoengineering, School of Electrical Engineering, Aalto University, P.O. Box 15500, 00076 Aalto, Finland
        \and 
            Zernike Institute for Advanced Materials, University of Groningen, Nijenborgh 4, 9747 AG, Groningen, The Netherlands
        \and
             Swedish Institute of Space Physics, P.O. Box 812, SE-981 28 Kiruna, Sweden
        \and
             Lule\aa{} University of Technology, Department of Computer Science, Electrical and Space Engineering, Kiruna,  SE-981 28, Sweden
        \and
            Science directorate, Chemistry \& Dynamics branch, NASA Langley Research Center, Hampton, VA 23666 Virginia, USA
        \and
            SSAI, Hampton, VA 23666 Virginia, USA    
        \and
            Royal Belgian Institute for Space Aeronomy, Avenue Circulaire 3, B-1180 Brussels, Belgium
        \and
            Department of Physics, Ume\aa{} University, 901 87 Ume\aa{}, Sweden
        \and 
            Department of Physics, Imperial College London, Prince Consort Road, London SW7 2AZ, United Kingdom
            }

   \date{\today}

  \abstract
   {Solar wind charge-changing reactions are of paramount importance to the physico-chemistry of the atmosphere of a comet, mass-loading the solar wind through an effective conversion of fast light solar wind ions into slow heavy cometary ions.}
   {To understand these processes and place them in the context of a solar wind plasma interacting with a neutral atmosphere, numerical or analytical models are necessary. Inputs of these models, such as collision cross sections and chemistry, are crucial.}
   {Book-keeping and fitting of experimentally measured charge-changing and ionization cross sections of hydrogen and helium particles in a water gas are discussed, with emphasis on the low-energy/low-velocity range that is characteristic of solar wind bulk speeds ($< 20$\,keV\,u$^{-1}$/$2000$\,\kms{}).
   }
   {We provide polynomial fits for cross sections of charge-changing and ionization reactions, and list the experimental needs for future studies. To take into account the energy distribution of the solar wind, we calculated Maxwellian-averaged cross sections and fitted them with bivariate polynomials for solar wind temperatures ranging from $10^5$ to $10^6$\,K ($12-130$\,eV).
   }
   {Single- and double-electron captures by He$^{2+}$ dominate at typical solar wind speeds. Correspondingly, single-electron capture by H$^+$ and single-electron loss by H$^-$ dominate at these speeds, resulting in the production of energetic neutral atoms (ENAs). Ionization cross sections all peak at energies above $20$\,keV and are expected to play a moderate role in the total ion production. However, the effect of solar wind Maxwellian temperatures is found to be maximum for cross sections peaking at higher energies, suggesting that local heating at shock structures in cometary and planetary environments may favor processes previously thought to be negligible. This study is the first part in a series of three on charge exchange and ionization processes at comets, with a specific application to comet 67P/Churyumov-Gerasimenko and the \emph{Rosetta} mission. 
   }


   \maketitle
%

\section{Introduction}

Over the past decades, evidence of charge-exchange reactions (CX) has been discovered in astrophysics environments, from cometary and planetary atmospheres to the heliosphere and to supernovae environments \citep{Dennerl2010}. They consist of the transfer of one or several electrons from the outer shells of neutral atoms or molecules, denoted M, to an impinging ion, noted X$^{i+}$, where $i$ is the initial charge number of species X. Electron capture of $q$ electrons takes the form
\begin{align}
        \textnormal{X}^{i+} + \textnormal{M} &\longrightarrow \textnormal{X}^{(i-q)+} + [\textnormal{M}]^{q+}. \label{eq:capture}
\end{align}

From the point of view of the impinging ion, a reverse charge-changing process is the electron loss (or stripping); starting from species $\textnormal{X}^{(i-q)+}$, it results in the emission of $q$ electrons:
\begin{align}
        \textnormal{X}^{(i-q)+} + \textnormal{M} &\longrightarrow \textnormal{X}^{i+} + [\textnormal{M}]\ +\ qe^-. \label{eq:loss}
\end{align}

For $q=1$, the processes are referred to as one-electron charge-changing reaction; for $q=2$, two-electron or double charge-changing reactions, and so on. The qualifier "charge-changing" encompasses both capture and stripping reactions, whereas "charge exchange" or "charge transfer" denote electron capture reactions only.
Moreover, "[M]" refers here to the possibility for compound M to undergo dissociation, excitation, and ionization, or a combination of these processes.  

Charge exchange was initially studied as a diagnostic for man-made plasmas \citep[][]{Isler1977,Hoekstra1998}. The discovery by \cite{Lisse1996} of X-ray emissions at comet Hyakutake C/1996 B2 was first explained by \cite{Cravens1997} as the result of charge-transfer reactions between highly charged solar wind oxygen ions and the cometary neutral atmosphere. Since this first discovery, cometary charge-exchange emission has successfully been used to remotely $(i)$ measure the speed of the solar wind \citep{Bodewits2004ApJ}, $(ii)$ measure its composition \citep{Kharchenko2003}, and thus the source region of the solar wind \citep{Bodewits2007AA,Schwadron2000}, $(iii)$ map plasma interaction structures \citep{Wegmann2005}, and more recently, $(iv)$ to determine the bulk composition of cometary atmospheres  \citep{Mullen2017}.

Observations of charge-exchanged helium, carbon and oxygen ions were made during the Giotto mission flyby of comet 1P/Halley and were reported by \cite{Fuselier1991}, who used a simplified continuity equation \citep[as in][]{Ip1989} to describe CX processes. \cite{Bodewits2004ApJ} reinterpreted their results with a new set of cross sections. More recently, the European Space Agency (ESA) \emph{Rosetta} mission to comet 67P/Churyumov-Gerasimenko (67P) between August 2014 and September 2016 provided a unique opportunity for studying CX processes in situ for an extended period of time \citep{Nilsson2015,CSW2016}. The observations need to be interpreted with the help of analytical and numerical models.

Charge state distributions and their evolution with respect to outgassing rate and cometocentric distance represent a proxy for the efficiency of charge-changing reactions at a comet such as 67P.
The accurate determination of relevant charge-changing and total ionization cross sections is a pivotal preliminary step when these reactions are to be quantified and in situ observations are to be interpreted.
Reviews of charge-changing cross sections exist, for example, for He$^{2+}$ particle electron capture cross sections in a variety of molecular and atomic target gases \citep{Hoekstra2006}, or for track-structure biological applications at relatively high energies \citep{Dingfelder2000,Uehara2002}. However, no critical and recent survey of charge-changing and ionization cross sections of helium and hydrogen particles in a water gas at solar wind energies is currently available. The goal of this paper is hence a critical review of experimental He and H charge-changing collisions with H$_2$O: in that, it complements the seminal study of \cite{Itikawa2005} for electron collisions with water by providing experiment-based datasets that space plasma modelers can easily implement, but also by assessing what future experimental work is needed.

In this study (Paper~I), we first discuss the method we used to critically evaluate CX and ionization cross sections. 
A review of existing experimental charge-changing and ionization cross sections of hydrogen and helium species in a water gas is then presented in Sections~\ref{sec:CXXsections} and \ref{sec:IonisationXsections}, with a specific emphasis on low-energy values for typical solar wind energies. As H$_2$O was the most abundant cometary neutral species during most of the \emph{Rosetta} mission \citep{Lauter2018}, we consider this species only. We identify laboratory data needs that are required to bridge the gaps in the existing experimental results. Polynomial fits for the systems $(\textnormal{H}^+,~\textnormal{H},~\textnormal{H}^-)-\textnormal{H}_2\textnormal{O}$ and $(\textnormal{He}^{2+},~\textnormal{He}^{+},~\textnormal{He})-\textnormal{H}_2\textnormal{O}$ are proposed. Recommended values are also tabulated for ease of book-keeping. In order to take into account the effect of the thermal energy distribution of the solar wind, Maxwellian-averaged charge-changing and ionization cross sections are discussed with respect to solar wind temperatures in Sect.~\ref{sec:maxwell}.

In a companion paper \citep[][hereafter Paper~II]{CSW2018b}, we then develop, based on these cross sections, an analytical model of solar wind charge-changing reactions in astrophysical environments, which we apply to solar wind-cometary atmosphere interactions. An interpretation of the \emph{Rosetta} ion and neutral datasets using this model is given in a separate iteration, namely \cite{CSW2018c}, hereafter Paper~III.


\section{Method}\label{sec:methodology}

We detail in this section the method we used in selecting cross sections. In this work, we only consider experimental inelastic (ionization and charge exchange) cross sections. Elastic (scattering) cross sections may play an important role at low impacting energies (a few tens of eV), leading to energy losses of the projectile species and to local heating. However, as shown in \cite{Behar2017}, solar wind ions, although highly deflected around the comet, do not display any significant slowing down at the position of \emph{Rosetta} in the inner coma: to a first approximation, elastic collisions may thus be neglected. 

Because H$_2$O was the main neutral species around comet 67P during the span of the \emph{Rosetta} mission, we only consider H$_2$O molecules as targets. However, it is important to remember that cometary environments contain other abundant molecules \citep[CO$_2$, CO, and O$_2$, see][]{Lauter2018}, and that parent molecules also photodissociate into H, O, C, H$_2$ , or OH fragments, which may in turn become dominant at very large cometocentric distances \citep[typically more than $100\,000$\,km for heliocentric distances below $2$\,AU, or astronomical units, see][]{Combi2004}. Because charge-transfer reactions are a cumulative process and depend on the column of atmosphere traversed \citep[see][]{CSW2016} and because some of these reactions may be resonant, their effect on the charge state distribution can potentially be large. Estimates of these effects using an analytical model of charge exchange at comets are discussed in Paper~II.

\subsection{Approach}
In selecting and choosing our chosen set of cross sections, our method consists of five steps:
\begin{itemize}
\item \textbf{Measurements} Survey of the currently published experimental cross sections $\sigma_{if}$ in H$_2$O vapor, with $i$ and $f$ the initial and final charge states of the projectile species considered. For example, $\sigma_{21}$ is the cross section of electron capture reaction $\textnormal{He}^{2+}\rightarrow\textnormal{He}^+$.
\item \textbf{Uncertainties}. Associated experimental uncertainties reported by the experimental teams. Sometimes, as in the case of \cite{Greenwood2004}, these uncertainties are statistical confidence intervals ($2\sigma$ standard deviation).
\item \textbf{Selection}. Selection of the chosen cross-section set, with emphasis on filling the low- and high-energy parts of the data. When experimental results are missing, we use the so-called additive rule (sometimes referred to as the "Bragg rule").
\item \textbf{Fit and validity}. Polynomial fits of the form
\begin{equation}
        \log_{10}(\sigma_{if}) = \sum_{j=0}^n{ p_j\, (\log_{10}\varv_i)^{j} }\end{equation}
are applied in a least-squares sense on the selected datasets as a function of impact speed $\varv_i$. Coefficients $p_j$ are the polynomial coefficients and $n$ is the degree of the polynomial fit. The degree of the fit is chosen so that in the energy range of the measurements and for every energy channel, fit residuals never exceed $15\%$ of the measurements.
A descriptive confidence level for the fit is also given, based on the agreement between the collected datasets and their respective datasets. It ranges from \emph{\textup{low}} ($>75\%$ uncertainty) to \emph{\textup{medium}} ($25-75\%$ uncertainty) and \textup{ \emph{\textup{high}}} ($<25\%$ uncertainty). 
Subscript $i$ in speeds and energies refers to "impactor" or "initial state", that is, the projectile speed or energy. 
\item \textbf{Further work}. We give recommendations on the necessary experimental work to be performed, and the energy range most critical to investigate.
\end{itemize}

\subsection{Extrapolations: the additive rule}
In several cases, we used the "additive rule" (that we refer to as AR in the following) to reconstruct missing H$_2$O datasets.
First expressed by \cite{Bragg1905} when investigating the stopping power of He$^{2+}$ in various atoms and molecules, it states that the stopping power of a molecule is, in a first approximation, equal to the sum of its individual atomic stopping powers. The AR hence assumes no intra-molecular effects, which leads to low predictability at energies where inelastic processes take place \citep{Thwaites1983}. For H$_2$O targets, this translates as
\begin{align}
    \sigma_{if}(\textnormal{H}_2\textnormal{O}) \sim 2\sigma_{if}(\textnormal{H}) + \sigma_{if}(\textnormal{O})
    \sim \sigma_{if}(\textnormal{H}_2) + \sigma_{if}(\textnormal{O}_2)/2.
\end{align}

At high impact energy, the AR for charge-changing cross sections has been well verified for protons and helium particles in many gases \citep{Toburen1968,Dagnac1970,Sataka1990,Endo2002}, both for electron capture \citep{Itoh1980capture} and for electron loss \citep{Itoh1980loss}. However, since this description is only empirical and not physical, one must be careful in applying it too systematically. For instance, it is well known that the AR breaks down for heavy ion collisions on complex molecules \citep{Wittkower1971,Bissinger1982}, for electron capture emission cross sections \citep{Bryan1990}, or at low energies \citep[see][]{Tolstikhina2018}.

In the case of low-energy extrapolations, the AR is not expected to be fulfilled because the molecular electrons move much faster than the projectile ion, and thus may follow the motion of the ion and adjust to it. Such an effect can be seen, for instance, in the low-energy electron capture cross-section measurements of \cite{Bodewits2006} on CO and CO$_2$ molecules, for which $\sigma_{21}$(CO)$>\sigma_{21}$(CO$_2$). When there were no experimental data, we used in this study the AR as an estimate for the cross sections at high energy and an indication of their magnitude at low energy, and always associated the retrieved cross sections with a high uncertainty.
When we applied the AR, we used the most recent experimental results for other species such as H$_2$, O$_2$ , or O and made a linear combination of their individual cross section to estimate that of H$_2$O. In several cases, when H$_2$O experimental results were available, the AR yielded results that are very different  (e.g., for $\sigma_{12}$ for the helium system, or for $\sigma_{01}$ and $\sigma_{0-1}$ for the hydrogen system), which lie typically within a multiplication factor $1-3$ of the H$_2$O results. In others, the AR is in good agreement (e.g., for $\sigma_{10}$ and $\sigma_{12}$ for the helium system, or, apparently, $\sigma_{-11}$ for the hydrogen system).
Consequently, when necessary and possible, we scaled the added cross sections to existing H$_2$O measurements to fill critical gaps in the datasets at either low or high energies.

Many charge-exchange and ionization cross sections for atoms and simple molecular targets are available as part of the charge-changing database maintained at the Lomonosov Moscow State University \citep{Novikov2009}. It is important to note that when available, cross sections for H$_2$ targets were preferred to those for H, in order to avoid resonant effects between protons and hydrogen atoms.

\subsection{Fitting of reconstructed cross sections}
Polynomial fits are here preferred to semi-empirical or more theoretical fits \citep{Dalgarno1958,Green1971} for their simplicity, versatility in describing the different processes, and standard implementation in complex physical models of cometary and astrophysical environments. Two broad categories of charge-exchange processes may take place: resonant (or symmetric) and non-resonant charge exchange \citep{Banks1973a}. Resonant charge exchange, such as $\textnormal{X}^+~+~\textnormal{X}~\rightarrow~\textnormal{X}~+~\textnormal{X}^+$, with ion $\textnormal{X}^+$ impacting its neutral counterpart X, usually has large cross sections; it has been shown theoretically that they continue to increase with decreasing impacting energies down to zero energy, where they peak \citep{Dalgarno1958}. For resonant capture at very high energies, where electron double-scattering dominates the interaction, \cite{Belkic1979} showed with theoretical considerations that the behavior of cross sections followed a $\varv^k$ power law, with $k = 11$. Conversely, non-resonant charge exchange peaks at non-zero velocity and is described by a more complex relation \citep{Lindsay2005}, with typical values at low (high) energies increasing (decreasing) as power laws of the velocity. We were able to use a simple polynomial fit of order $2-6$ to describe all charge-changing and ionization cross sections, which makes it easy to compare between them. The validity range of the fit was confined to the velocity range of available measurements. Where needed, smooth extrapolations of the fits were performed in power laws of the velocity down to $100$\,\kms{} and for very high energies; these extrapolations have large uncertainties and are only given for reference in the tables in the appendix.

We also note that in a cometary environment, resonant charge-exchange reactions such as $\textnormal{H}^+-\textnormal{H}$ may take place \citep{Bodewits2004ApJ}. For example, H and O are both present in the solar wind and in the cometary coma; at large cometocentric distances, cometary H and O atoms dominate the neutral coma because H$_2$O, CO$_2$ or CO will be fully photodissociated. Moreover, resonant processes usually have large cross sections. However, for a relatively low-activity comet such as comet 67P (outgassing rate lower than $10^{28}$\,s$^{-1}$), and although the hydrogen cometo-corona extends millions of kilometers upstream, the solar wind proton densities will have diminished due to resonant charge exchange by less than $1\%$ by the time it reaches a cometocentric distances of $10\,000$\,km. This point is further discussed in Paper~II.

\section{Experimental charge-changing cross sections for (H,~He) particles in H$_2$O}\label{sec:CXXsections}

Cross sections are given at typical solar wind speeds and are discussed in light of available laboratory measurements. Twelve cross sections, six listed in Sect.~\ref{sec:crs_helium} for helium and six in Sect.~\ref{sec:crs_hydrogen} for hydrogen, are considered.

Starting with an incoming ion species $\textnormal{X}^i$ in an initial charge state $i$ colliding with neutral target $\textnormal{M}$, and three possible final charge states $(i,~i-1,~i-2)$, the reactions can be written as

\begin{align*}
        \sigma_{i,i-1}:&\ \textnormal{X}^{i+} &+&\ \textnormal{M} &\longrightarrow&\ \textnormal{X}^{(i-1)+} &+&\ \textnormal{M}^+ &\ \textnormal{single capture} \\
    \sigma_{i,i-2}:&\ \textnormal{X}^{i+} &+&\ \textnormal{M} &\longrightarrow&\ \textnormal{X}^{(i-2)+} &+&\ \textnormal{M}^{2+} &\ \textnormal{double capture}\\    
    \sigma_{i-1,i}:&\ \textnormal{X}^{(i-1)+}  &+&\ \textnormal{M} &\longrightarrow&\ \textnormal{X}^{i+} &+&\ \textnormal{M} + e^- &\ \textnormal{single stripping} \\
    \sigma_{i-1,i-2}:&\ \textnormal{X}^{(i-1)+} &+&\ \textnormal{M} &\longrightarrow&\ \textnormal{X}^{(i-2)+} &+&\ \textnormal{M}^+ &\ \textnormal{single capture} \\
    \sigma_{i-2,i}:&\ \textnormal{X}^{(i-2)+} &+&\ \textnormal{M} &\longrightarrow&\ \textnormal{X}^{i+} &+&\ \textnormal{M} + 2e^- &\ \textnormal{double stripping} \\
    \sigma_{i-2,i-1}:&\ \textnormal{X}^{(i-2)+} &+&\ \textnormal{M} &\longrightarrow&\ \textnormal{X}^{(i-1)+} &+&\ \textnormal{M} + e^- &\ \textnormal{single stripping.}
\end{align*}

\begin{figure}
  \includegraphics[width=\linewidth]{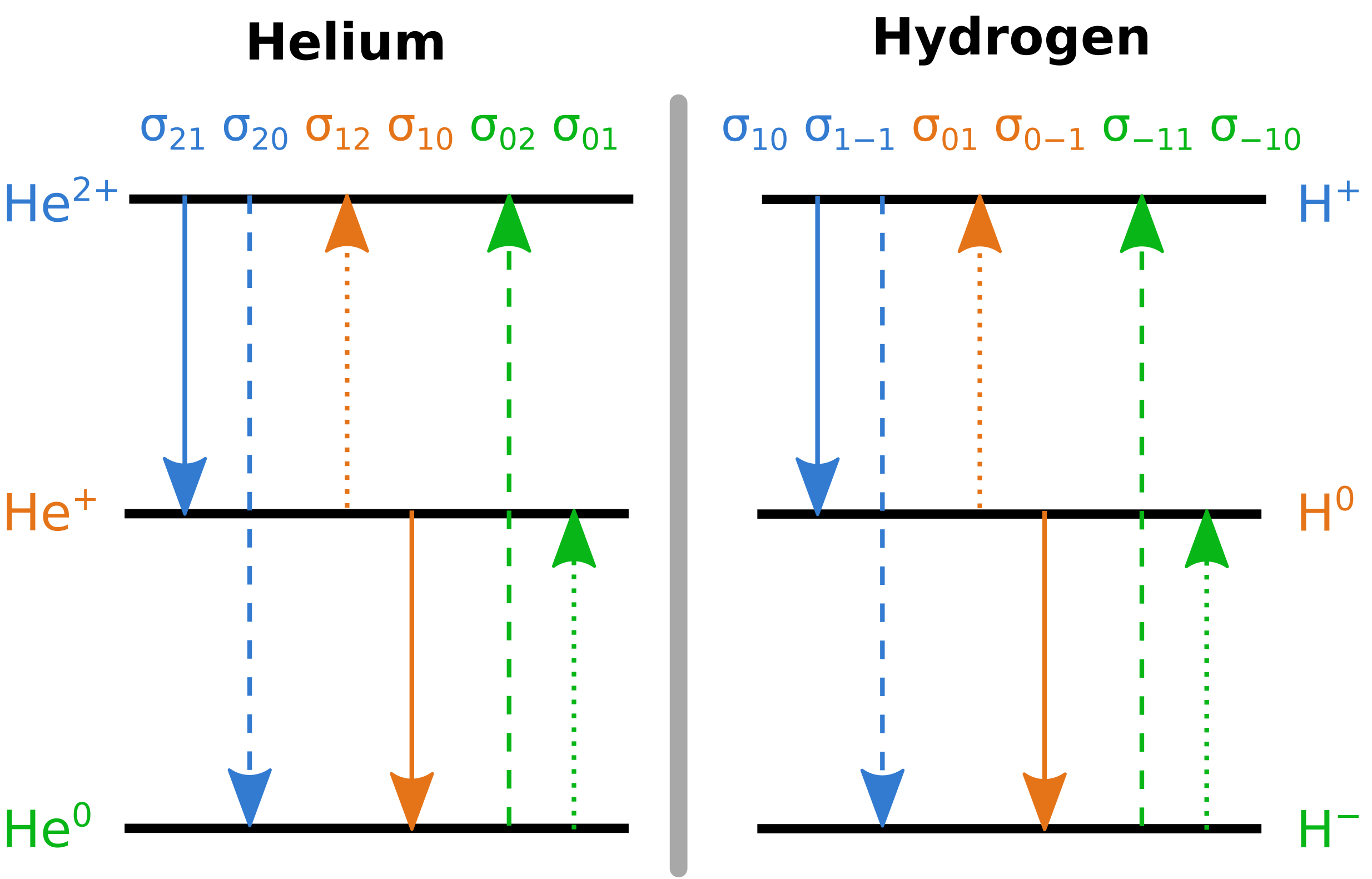}
     \caption{Charge-changing reactions for helium and hydrogen in a gas. One-electron capture processes are depicted with a solid line, one-electron loss processes with a dotted line, and double charge-changing reactions with a dashed line. Cross sections $\sigma_{ij}$ from initial charged state $i$ to final charged state $j$ are indicated.}
 \label{fig:CXLevels}
 \end{figure}

Figure~\ref{fig:CXLevels} illustrates the six processes per impacting species (hydrogen, initial charge states $i=1, 0, -1$, and helium, initial charge states $i=2, 1, 0$), with the chosen nomenclature for the charge-changing cross sections.

A molecular target such as H$_2$O may dissociate into atomic or molecular fragments through electron capture or stripping \citep[see][in H$^+$ and He$^{2+}$-H$_2$O collisions]{Luna2007,Alvarado2005}; similarly, the impacting species may become excited in the process \citep[see][in He$^{2+}-$H$_2$O collisions]{Seredyuk2005}. For the remainder of this paper, only total charge-changing cross sections are considered, that is, the sum of all dissociation and excitation channels. In other words, we only consider the loss of solar wind ions, not the production of excited or dissociated ionospheric species. 

\subsection{Helium projectiles}\label{sec:crs_helium}

The helium projectiles we considered are He$^{2+}$, He$^+$ and He$^0$. Charge-changing cross sections for H$_2$O are presented, and our choice for each cross section is given.
We note that all impact energies for helium are quoted in keV per amu (abbreviated keV/u), 
allowing us to compare the results of different experiments where sometimes $^3$He isotopes are used instead of the more common $^4$He. 

\begin{figure*}
  \includegraphics[width=\linewidth]{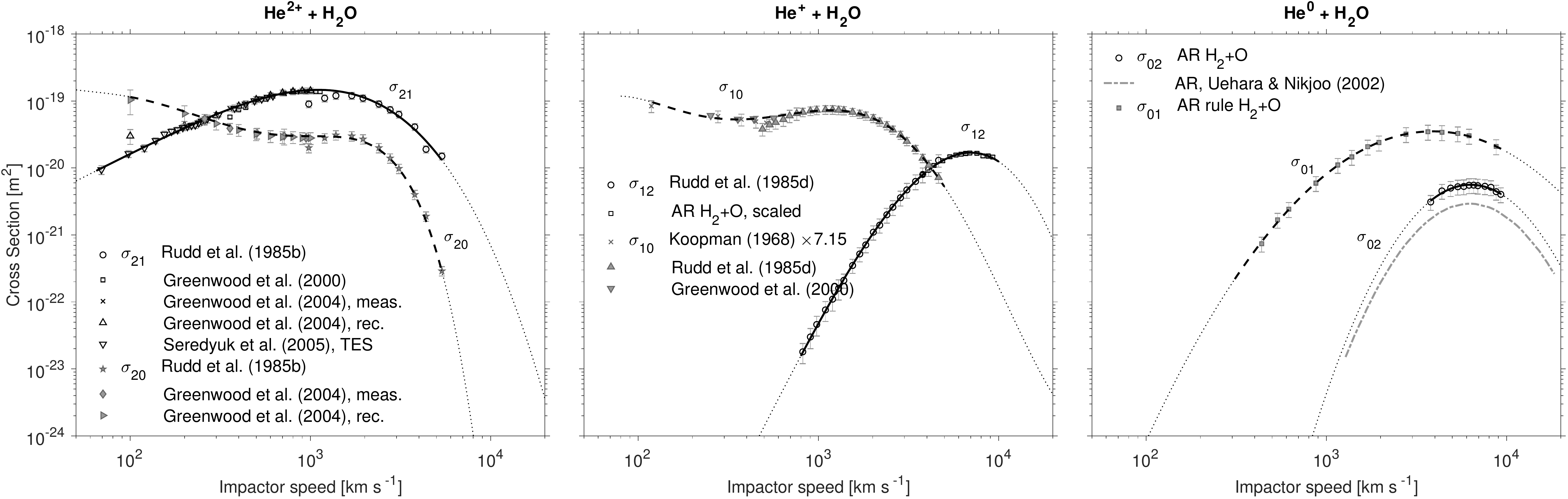}
     \caption{Experimental charge-changing cross sections for fast helium atoms and ions in a water gas as a function of impact speed.  "AR" refers to the "additive rule": when no experimental results for H$_2$O are available, results for H$_2$ and O$_2$ are combined to give an estimate (see text for details); experimental uncertainties for these estimates are at least $25$\%. Recommended polynomial fits in thick continuous or dashed lines are also shown, whose coefficients are listed in Table~\ref{tab:heliumXsections}. Smooth extrapolations at low and high energies are indicated as thin dotted lines.
    }
     \label{fig:crossSection_CX_He}
 \end{figure*}

Cross sections and their corresponding recommended fits are plotted in Fig.~\ref{fig:crossSection_CX_He}. Polynomial fitting coefficients are listed in Table~\ref{tab:heliumXsections}.

  \begin{table*}
        \centering
        \caption{Recommended charge-changing cross section polynomial fits for (He$^{2+}$, He$^+$, and He$^0$) projectiles colliding with H$_2$O vapor.}
        \label{tab:heliumXsections}
    \tiny
        \begin{tabular}{lcrrrrrrllc} 
                \hline\hline
        Cross section & Degree &  \multicolumn{6}{c}{Coefficients}  & \multicolumn{2}{c}{Validity range} & Confidence\\ \relax
                [m$^{2}$] & $n$ & $p_0$ & $p_1$ & $p_2$ & $p_3$ & $p_4$ & $p_5$ & $\varv_i$ [\kms{}] & $E_i$ [keV/u] & \\
                \hline
        $\sigma_{21}$ & $4$ & $-129.6349$   & $85.3069$ & $-25.7100$  & $3.50927$ & $-0.18016$ & - & $75-5350$ & $0.03-150$ & high\\
                $\sigma_{20}$ &  $5$ & $3327.3456$ & $-3277.1313$ & $1272.0445$ & $-244.7593$ & $23.35760$ & $-0.88492$ & $100-5350$ & $0.05-150$ & high\\
                $\sigma_{12}$ & $4$ & $-314.9414$ & $205.4565$ & $-57.4465$ & $ 7.3111$ & $-0.34819$ & - & $820-10\,000$ & $3.50-520$ & high\\ 
                $\sigma_{10}$ & $5$ & $-5450.8180$ & $4667.4695$ & $-1592.4968$ & $269.5820$ & $-22.63262$ & $0.75346$ & $120-5000$ & $0.08-130$ & high\\
                $\sigma_{02}$ & $2$ & $-245.4003$ & $66.2165$ & $-4.8686$ & - & - & - & $3800-9300$ & $75.0-450$ & low\\  
                $\sigma_{01}$ & $2$ & $-98.7467$ & $24.0604$ & $-1.8252$ & - & - & - & $310-10\,000$ & $0.50-520$ & low\\   
                \hline
        \end{tabular}
        \tablefoot{The polynomial, function of the speed of the impactor, is of the form $\log_{10}(\sigma) = \sum_{j=0}^n{p_j\,(\log_{10} \varv_i)^{j}}$, where $n$ is the degree of the fit, the speed $\varv_i$ is expressed in m\,s$^{-1}$ , and the cross section $\sigma$ in m$^2$. Ranges of validity for impact speeds and energies are given. Confidence levels on the fits are indicated: high ($<25\%$), medium ($25-75\%$), and low ($>75\%$).}
 \end{table*}

\subsubsection{He$^{2+}$ -- H$_2$O reactions}

Reactions involving He$^{2+}$ are the one-electron $\sigma_{21}$ and two-electron $\sigma_{20}$ captures. They are shown in Fig.~\ref{fig:crossSection_CX_He} (left).

\begin{itemize}[label=$\bullet$]
    \item Reaction $\sigma_{21}$ ($\textnormal{He}^{2+}\rightarrow\textnormal{He}^+$)

    \begin{itemize}
            \item \textbf{Measurements.}  Measurements of the one-electron capture by He$^{2+}$ in a water gas were reported by \cite{Greenwood2004} in the $0.35-4.67$\,keV/u energy range and by \cite{Rudd1985_Hepp} for $E_i = 5-150$\,keV/u (for $^3$He isotopes). \cite{Greenwood2000} also made measurements up to $6.67$\,keV/u): their values are in excellent agreement with the subsequent results from the same team, except at $0.67$\,keV/u ($\varv_i = 360$\,\kms{}), where it is about $25$\% smaller. We note that \cite{Greenwood2004} provide recommended values that extend the valid range to $0.052-5.19$\,keV/u ($100-1000$\,\kms{}). At $5$ and $7.5$\,keV/u, \cite{Rudd1985_Hepp} appear to underestimate the cross section by about $35$\% with respect to that measured by \cite{Greenwood2000}. \cite{Seredyuk2005} and \cite{Bodewits2006} measured state-selective charge-exchange cross sections between $0.025$\,keV/u and $12$\,keV using two complementary techniques (fragment ion spectroscopy, and translational energy spectrometry, or TES): below $0.25$\,keV/u, capture into the He$^+ (n=1)$ state dominates, whereas capture into the He$^+ (n=2)$ state is dominant above this energy. Their total TES cross-section results were normalized to those of \cite{Greenwood2004}, and display a matching energy-dependence with respect to the reference measurements.
        \item \textbf{Uncertainties.} On average, uncertainties are about $10$\% at low energies \citep{Greenwood2004,Seredyuk2005} ($15-25$\% below $0.3$\,keV/u, $95\%$ confidence interval) and $12$\% at high energies \citep{Rudd1985_Hepp}. 
        \item \textbf{Selection.} All datasets connect rather well at their common limit, if we discard the \cite{Rudd1985_Hepp} measurements below $8$\,keV/u. We chose to use the values of \cite{Seredyuk2005} between $0.025-2$\,keV/u, those of \cite{Greenwood2004} between $2-5.19$\,keV/u supplemented up to $6.67$\,keV/u by those of \cite{Greenwood2000}, and we extend the set to energies above $10$\,keV/u with those of \cite{Rudd1985_Hepp}.
        \item \textbf{Fit and validity.} A least-squares polynomial fit of degree $3$ in $\log_{10}$ of the He$^{2+}$ speed $\varv_i$ was performed. Expected validity range $\varv_i = 75-5350$ \kms{} ($E_i = 0.03-150$\,keV/u). Confidence: high. 
        \item \textbf{Further work.} Need for very low-energy measurements, that is, for $E_i < 0.02$\,keV/u. 
    \end{itemize}

\item Reaction $\sigma_{20}$  ($\textnormal{He}^{2+}\rightarrow\textnormal{He}^0$)
        \begin{itemize}
        \item  \textbf{Measurements.}  Cross sections for the two-electron capture by He$^{2+}$ from water vapor were experimentally measured by \cite{Greenwood2004} for $E_i = 0.35-4.67$\,keV/u and by \cite{Rudd1985_Hepp} between $5$ and $150$\,keV/u. As for $\sigma_{21}$, \cite{Greenwood2004} gave fitted recommendations, extending their dataset to $0.052-5.19$\,keV/u. 
        \item \textbf{Uncertainties.} Uncertainties range on average between $20$\% below $5$\,keV/u ($30-40$\% below $0.3$\,keV/u) \citep{Greenwood2004} to $16$\% above it \citep{Rudd1985_Hepp}. 
        \item \textbf{Selection.} Although as previously, \cite{Rudd1985_Hepp} do seem to underestimate the cross section at $5$\,keV/u by about $30$\%, both datasets join together well if we discard this first data point. We chose to use the \cite{Greenwood2004} recommendation for $E_i = 0.052-5.19$\,keV/u and \cite{Rudd1985_Hepp} for $E_i>5$\,keV/u. 
        \item \textbf{Fit and validity.} A polynomial fit of order $5$ best represents the datasets. Expected validity range: $\varv_i=100-5350$\,\kms{} ($E_i=0.05-150$\,keV/u). Confidence: high. 
        \item \textbf{Further work.} Need for very low-energy measurements, that is, for $E_i<0.1$\,keV/u.
    \end{itemize}
\end{itemize}

\subsubsection{He$^{+}$ -- H$_2$O reactions}
Reactions involving He$^+$ ions are the one-electron loss $\sigma_{12}$ and the one-electron capture $\sigma_{10}$. They are shown in Fig.~\ref{fig:crossSection_CX_He} (middle).

\begin{itemize}[label=$\bullet$]
\item Reaction $\sigma_{12}$ ($\textnormal{He}^{+}\rightarrow\textnormal{He}^{2+}$)
    \begin{itemize}
        \item \textbf{Measurements.} \cite{Rudd1985_Hep} measured the one-electron loss cross section for He$^+$ in water in the $3.50-112.5$\,keV/u ($820-4640$\,\kms{}) range. No measurements are available below or above these energies. 
            \item \textbf{Uncertainties.} Uncertainties are $21-33$\% on average \citep{Rudd1985_Hep}. 
            \item \textbf{Selection.} We chose to use the measurements by \cite{Rudd1985_Hep}, and following the recommendation of \cite{Uehara2002}, we used the additive rule with the cross sections of \cite{Sataka1990} in H$_2$ and O$_2$ at energies between $75$ and $450$\,keV/u to define the peak of the cross section at high energies. At overlapping energies, the reconstructed cross section $\sigma(\textnormal{H}_2)+\sigma(\textnormal{O}_2)/2$ is lower than that measured by \cite{Rudd1985_Hep} in H$_2$O: the latter measurements at $75$\,keV/u were used to calibrate the former, resulting in a constant multiplication factor of $1.64$ for the H$_2$O dataset at high energies reconstructed from \cite{Sataka1990}. 
            \item \textbf{Fit and validity.} A polynomial fit in $\log_{10}$ of order $4$ was used. Validity range: $\varv_i=820-10\,000$\,\kms{} ($E_i=3.5-520$\,keV/u). Confidence: high.   
            \textbf{Further work.} Need for low- ($0.01<E_i<5$\,keV/u) and high-energy ($E_i>100$\,keV/u) measurements.
    \end{itemize}

\item Reaction $\sigma_{10}$ ($\textnormal{He}^{+}\rightarrow\textnormal{He}^{0}$)
    \begin{itemize}
        \item \textbf{Measurements.}  Measurements of the one-electron capture cross section of fast He$^+$ ions in water were made by \cite{Koopman1968} between $0.2$ and $1.4$\,keV/u energy, \citep{Rudd1985_Hep} in the $1.25-112.5$\,keV/u ($490-4640$\,\kms{}) range and by \cite{Greenwood2000} for $0.3-1.7$\,keV/u ($253-565$\,\kms{}). The results reported by \cite{Koopman1968} are a factor $7.15$ lower than those of \cite{Greenwood2000} at their closest common energy ($0.33$\,keV/u), but are nonetheless qualitatively similar in shape and energy behavior. 
            \item \textbf{Uncertainties.} Uncertainties are below $7$\% for $E_i<1.7$\,keV/u \citep{Greenwood2000} and span $14-20$\% for $E_i >2$\,keV/u \citep{Rudd1985_Hep}. \cite{Koopman1968} claimed an uncertainty of $20$\%. 
            \item \textbf{Selection.} The three datasets significantly differ in their common energy range \citep[$>30$\%, to almost an order of magnitude for][]{Koopman1968}. Because the \cite{Greenwood2000} measurements have a higher accuracy, we chose this dataset below $1.7$\,keV/u and used \cite{Rudd1985_Hep}'s for  $E_i\geq2.5$\,keV/u. As remarked by \cite{Koopman1968}, the cross section is expected to continue to rise with diminishing energies, which may be due to a near-resonant process involving highly excited states of H$_2$O$^+$. This tendency is also seen with electron capture by He$^+$ impinging on a O$_2$ gas \citep{Mahadevan1968}. We therefore supplemented our data at low energy with an adjustment of the \cite{Koopman1968} measurement at $73$\,eV/u ($118$\,\kms{}) by multiplying by a calibrating factor of $7.15$ ($\sigma_{10}^\textnormal{adj} \approx 9\times10^{-20}$\,m$^2$), and placing less weight on this particular dataset because of the large uncertainties. We note that the additive rule using the results of \cite{Rudd1985_Hep_gases} for H$_2$ and O$_2$ agress well with the measurements made in H$_2$O (within the experimental uncertainties). 
            \item \textbf{Fit and validity.} Polynomial fit of order $5$ was performed. Validity range: $\varv_i=120-5000$\,\kms{} ($E_i = 0.08-130$\,keV/u). Confidence: high. 
            \item \textbf{Further work.} Need for measurements in the very low-energy range, that is, $E_i<0.5$\,keV/u.
    \end{itemize}
\end{itemize}

\subsubsection{He$^{0}$ -- H$_2$O reactions}
The reactions involving the neutral atom He$^0$ are the two-electron $\sigma_{02}$ and  one-electron $\sigma_{01}$ losses. They are shown in Fig.~\ref{fig:crossSection_CX_He} (right).

\begin{itemize}[label=$\bullet$]
\item Reaction $\sigma_{02}$ ($\textnormal{He}^{0}\rightarrow\textnormal{He}^{2+}$)
    \begin{itemize}
        \item \textbf{Measurements.} No measurement of the two-electron loss cross section for helium atoms in a water gas has been reported.  
            \item \textbf{Uncertainties.} N/A.  
            \item \textbf{Selection.} Because of the lack of measurements, we chose to use the additive rule so that $\sigma_{02}$(H$_2$O)$\sim\sigma_{02}$(H$_2$)+$\sigma_{02}$(O$_2)/2$. For H$_2$ and O$_2$, and following \cite{Uehara2002}, we used the measurements of \cite{Sataka1990} ($75-450$\,keV/u), which were performed around the cross-section peak with an uncertainty below $7$\%. The composite fit of \cite{Uehara2002} is within a factor~$2$ and extends down in energies to about $8.5$\,keV.
            \item \textbf{Fit and validity.} A polynomial fit of order $2$ was performed. Validity range: $\varv_i=3800-9300$\,\kms{} ($E_i=75-450$\,keV/u). Confidence: low. 
            \item \textbf{Further work.} Need of measurements at any energy, with priority for $0.05<E_i<500$\,keV/u.
    \end{itemize}
    
\item Reaction $\sigma_{01}$ ($\textnormal{He}^{0}\rightarrow\textnormal{He}^{+}$)
    \begin{itemize}
        \item \textbf{Measurements.} No measurement of the one-electron loss cross section for helium atoms in a water gas has been reported.   
            \item \textbf{Uncertainties.} N/A. 
            \item \textbf{Selection.} Because of  the lack of measurements, we chose to use the additive rule so that $\sigma_{01}$(H$_2$O)$\sim\sigma_{01}$(H$_2$)+$\sigma_{01}$(O$_2)/2$. For H$_2$, we used the recommendation of \cite{Barnett1990} (who analyzed all measurements prior to 1990) in the $0.5-10^3$\,keV/u energy range and supplemented them by the more recent measurements of \cite{Sataka1990} ($75-450$\,keV/u), which are both in excellent agreement. For O$_2$, we used the results of \cite{Allison1958} between $1$ and $50$\,keV/u and \cite{Sataka1990} between $75$ and $450$\,keV/u; these datasets connect very well around $60$\,keV/u. Associated uncertainties of separate cross sections are better than $10$\%. 
            \item \textbf{Fit and validity.} A polynomial fit of order $2$ was performed. Validity range: $\varv_i=310-10\,000$\,\kms{} ($E_i = 0.50-520$\,keV/u). Confidence: low. 
            \item \textbf{Further work.} Need of measurements at any energy, with priority for $0.05<E_i<500$\,keV/u.
    \end{itemize}
\end{itemize}

\subsubsection{Discussion}
Figure~\ref{fig:crossSection_CX_He} shows that all charge-changing cross sections peak at values around $10^{-19}-10^{-20}$\,m$^2$. Except for the capture cross sections $\sigma_{20}$ (He$^{2+}\rightarrow$He) 
 and $\sigma_{10}$ (He$^{+}\rightarrow$He), which display a peak at speeds below $100$\,\kms{}, the main peak of all other cross sections is situated at speeds higher than $1000$\,\kms{}. $\sigma_{21}$ (He$^{2+}\rightarrow$He$^{+}$) is the largest cross section between $400$ and $3500$\,\kms{} (peak at $1.5\times10^{-19}$\,m$^{2}$), whereas at low speeds, both double- and single-electron captures $\sigma_{20}$ and $\sigma_{10}$ for He$^{2+}$ and He$^+$ impactors become dominant, reaching values of about $1\times10^{-19}$\,m$^{2}$ at $100$\,\kms{}.  
 Comparatively, the electron-loss cross sections from atomic He and from He$^+$ start to become significant at speeds above $3000$\,\kms{}, where they reach a maximum and where electron capture cross sections start to decrease. The largest of these cross sections, stripping cross section $\sigma_{01}$, reaches values of $3.5\times10^{-20}$\,m$^2$ at its peak.

\subsection{Hydrogen projectiles}\label{sec:crs_hydrogen}
The hydrogen projectiles we considered are H$^{+}$, H$^0$ , and H$^-$. Charge-changing cross sections for H$_2$O are presented, and our choice for each cross section is given, following the template of Sect.~\ref{sec:crs_helium}.

\begin{figure*}
  \includegraphics[width=\linewidth]{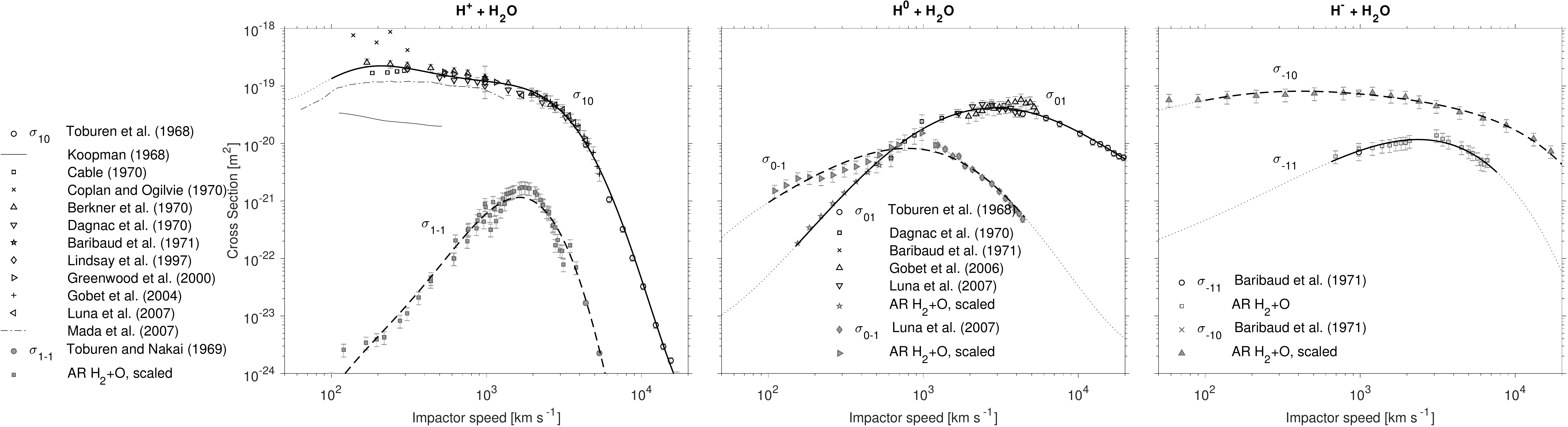}
     \caption{Experimental charge-changing cross sections for fast hydrogen atoms and ions in a water gas as a function of impact speed. "AR" refers to the "additive rule": when no experimental results for H$_2$O are available, results for H$_2$ and O$_2$ are combined to give an estimate (see text for details); experimental uncertainties for these estimates are at least $25$\%. Polynomial fits in thick continuous or dashed lines are also shown, whose coefficients are listed in Table~\ref{tab:hydrogenXsections}. Smooth extrapolations at low and high energies are indicated as thin dotted lines. 
     }
     \label{fig:crossSection_CX_H}
 \end{figure*}

The cross sections and their corresponding recommended fits are plotted in Fig.~\ref{fig:crossSection_CX_H}. Polynomial fit coefficients are listed in Table~\ref{tab:hydrogenXsections}.

 \begin{table*}
        \centering
        \caption{Recommended charge-changing cross-section polynomial fits for (H$^{+}$, H$^0$, H$^-$) projectiles colliding with H$_2$O vapor.}
        \label{tab:hydrogenXsections}
    \scriptsize
        \begin{tabular}{lcrrrrrrrllc} 
                \hline\hline
        Cross section & Degree &  \multicolumn{7}{c}{Coefficients}  & \multicolumn{2}{c}{Validity range} & Confidence\\ \relax
                [m$^{2}$] & $n$ & $p_0$ & $p_1$ & $p_2$ & $p_3$ & $p_4$ & $p_5$ & $p_6$ & $\varv_i$ [\kms{}] & $E_i$ [keV/u] & \\
                \hline
        $\sigma_{10}$ & $6$ & $33151.6652$ & $-34755.9561$ & $15090.7461$ & $-3475.4246$ & $447.74484$ & $-30.59282$ & $0.865965$ & $100-20\,000$ & $0.05-2100$ & high\\
                $\sigma_{1-1}$ & $6$ & $23065.9322$ & $-25763.6624$ & $11868.1014$ & $-2890.6324$ & $392.69112$ & $-28.20982$ & $0.837012$ & $100-7600$ & $0.05-300$ & low\\  
                $\sigma_{01}$ & $4$ & $332.2201$ & $ -247.4856$ & $62.7381$ & $-6.8495$ & $0.27305$ & - & - & $150-20\,000$ & $0.10-2100$ & medium \\
                $\sigma_{0-1}$ & $5$ & $-1446.6760$ & $1267.4240$ & $-450.7229$ & $79.8827$ & $-7.03258$ & $0.24532$ & - & $100-4500$ & $0.05-105$ & low\\ 
                $\sigma_{-11}$ & $4$ & $-121.2559$ & $79.1328$ & $-23.9740$ & $3.2444$ & $-0.16311$ & - & - & $650-7600$ & $2.20-300$ & low\\ 
                $\sigma_{-10}$ & $5$ & $403.3783$ & $-383.3741$ & $137.3481$ & $-24.3561$ & $2.14359$ & $-0.07510$ & - & $100-20\,000$ & $0.05-2100$ & low\\ 
                \hline
        \end{tabular}
        \tablefoot{The polynomial, function of the speed of the impactor, is of the form $\log_{10}(\sigma) = \sum_{j=0}^n{p_j\,(\log_{10} \varv_i)^{j}}$, where $n$ is the degree of the fit, the speed $\varv_i$ is expressed in m\,s$^{-1}$ and the cross sections $\sigma$ in m$^2$. Ranges of validity for impact speeds and energies are given. Confidence levels on the fits are indicated as high ($<25\%$), medium ($25-75\%$), and low ($>75\%$) (see text).}
 \end{table*}

\subsubsection{H$^+$--H$_2$O}

Reactions involving H$^+$ are the one-electron $\sigma_{10}$ and two-electron $\sigma_{1-1}$ captures. They are shown in Fig.~\ref{fig:crossSection_CX_H} (left). 
\begin{itemize}[label=$\bullet$]
    \item Reaction $\sigma_{10}$ ($\textnormal{H}^{+}\rightarrow\textnormal{H}^{0}$)
    \begin{itemize}
            \item \textbf{Measurements.} Since the end of the 1960s, many investigators have measured the one-electron capture cross section for protons in water \citep{Koopman1968,Toburen1968,Berkner1970,Cable1970,Coplan1970,Dagnac1970,Rudd1985_Hp_H2O}, concentrating on relatively high impact energies \citep[$E_i>1$\,keV, see][]{Barnett1977}. Recently, the cross section was remeasured by \cite{Lindsay1997} ($E_i = 0.5-1.5$\,keV) and by \cite{Greenwood2000} ($1.5-7$\,keV). At high energies ($15<E_i<150$\,keV), the recent measurements of \cite{Gobet2004} and \cite{Luna2007} agree well with those of \cite{Toburen1968}. All measurements are in excellent agreement, except for those by \cite{Coplan1970}, who seemed to overestimate their results by a factor $2-4$, and \cite{Koopman1968}, who underestimate them by about one order of magnitude. Finally, \cite{Baribaud1971} and \cite{Baribaud1972} reported a value of $\sigma_{10} = (14\pm8)\times10^{-20}$\,m$^2$ at $5$\,keV, in good agreement with the other measurements. It is interesting to remark that the additive rule estimates using data in H$_2$ \citep{Gealy1987_Hp} and O \citep{VanZyl2014} are $30$\% lower on average than the direct measurements in H$_2$O. 
        \item \textbf{Uncertainties.} Measurement errors for the recent datasets are smaller than $10$\% on average \citep{Lindsay1997,Greenwood2000}. 
        \item \textbf{Selection.} To extrapolate at energies below $500$\,eV with a plausible energy dependence, we used the theoretical calculations of \cite{Mada2007} (Fig.~6, total charge-transfer cross section including all molecular axis collision orientations) increased by a factor $2.2$ to match \cite{Greenwood2000} and \cite{Lindsay1997} at $500$\,eV. At high energies, the results of \cite{Luna2007}, combined with those of \cite{Gobet2004}, were chosen. 
        \item \textbf{Fit and validity.} A least-squares polynomial fit of degree $6$ in $\log_{10}$ of the proton speed $\varv_i$ was performed.
        Expected validity range $\varv_i = 100-2\times10^4$ \kms{} ($E_i = 0.05-2100$\,keV). Confidence: high. 
        \item \textbf{Further work.} Measurements in the low-energy range $0.05~<E_i<5$\,keV with good energy resolution are needed. 
    \end{itemize}
    
        \item Reaction $\sigma_{1-1}$ ($\textnormal{H}^{+}\rightarrow\textnormal{H}^{-}$) 
        \begin{itemize}
            \item \textbf{Measurements.} Only one measurement of the double-electron capture by protons in H$_2$O has been reported \citep{Toburen1969}, and at high energies ($75<E_i<250$\,keV). No low-energy measurements are available. 
        \item \textbf{Uncertainties.} Errors are reported to be $8$\% in this high energy range. 
        \item \textbf{Selection.} Lacking data, we used the additive rule for the double capture by H$_2$ , which is well documented \citep{Allison1958,McClure1963,Kozlov1966,Williams1966DC,Schryber1967,Toburen1969,Salazar2009}, and O$_2$ \citep[][given per atom of oxygen]{Allison1958} at low proton impact energies. We supplement these estimates with the measurements in water by \cite{Toburen1969} at high energies. Since the measurements reported by \cite{Allison1958} for O$_2$ are only made around $10$\,keV, the behavior of H$_2$O at energies below is unknown. We chose to reconstruct the H$_2$O data around the peak with the additive rule and to multiply the H$_2$+O data at low energies by a factor $\sigma_{\textnormal{H}_2\textnormal{O}}/\sigma_{\textnormal{H}_2} = 3.3$ to connect smoothly with the peak H$_2$O cross section. $1\sigma$ uncertainties for the AR dataset are indicated in the figure.
        \item \textbf{Fit and validity.} A polynomial fit of order $6$ in $\log_{10}$ was performed on the overall reconstructed cross section. Because of the reconstructed AR dataset, the fit underestimates the cross-section peak by about $50\%$, although uncertainties are likely much larger. Validity range $\varv_i = 100-7600$ \kms{} ($E_i = 0.05-300$\,keV). Confidence: low. 
        \item \textbf{Further work.} Need of measurements for $0.05<E_i<100$\,keV to confirm this estimate.    
    \end{itemize}
\end{itemize}

\subsubsection{H$^0$ -- H$_2$O}
Reactions involving H$^0$ are the one-electron loss $\sigma_{01}$ and the one-electron capture $\sigma_{0-1}$. They are shown in Fig.~\ref{fig:crossSection_CX_H} (middle). 

\begin{itemize}[label=$\bullet$]
        \item Reaction $\sigma_{01}$ ($\textnormal{H}^{0}\rightarrow\textnormal{H}^{+}$) 
        \begin{itemize}
            \item \textbf{Measurements.} \cite{Dagnac1969,Dagnac1970} measured one-electron-loss cross sections for the hydrogen impact on H$_2$O between $1.5$ and $60$\,keV, which are in excellent agreement in their common range with the newer values given by \cite{Luna2007} in the $15-90$\,keV range, which include both reaction channels $\textnormal{H} \rightarrow \textnormal{H}^+ + \textnormal{H}_2\textnormal{O}^{+} + 2e$ and $\rightarrow \textnormal{H}^+ + \textnormal{H}_2\textnormal{O} + e$. \cite{Baribaud1971} and \cite{Baribaud1972} reported a value of $\sigma_{01} = (1.6\pm0.8)\times10^{-20}$\,m$^2$ at $5$\,keV in good agreement. \cite{Gobet2006} reported cross sections between $20$ and $150$\,keV, whereas \cite{Toburen1968} made measurements between $100$\,keV and $2500$\,keV, all in excellent agreement. 
        \item \textbf{Uncertainties.} Uncertainties range from 30\% ($1.5-5$\,keV) to $12-15$\% ($>5$\,keV) \citep{Dagnac1970,Luna2007} and are on the order of $25$\% at very high energies \citep{Gobet2006}. 
        \item \textbf{Selection.} We used data from \cite{Dagnac1970} and \cite{Luna2007} between $1.5$ and $90$\,keV. To extrapolate the behavior of the cross section at lower energies, we used the additive rule $\sigma_{01}$(H$_2$)+$\sigma_{01}$(O), using \cite{Gealy1987_H} for H impact on H$_2$ paired with data reported by \cite{VanZyl2014} for H impact on O (both with uncertainties of about $15-25$\%) between $0.125$ and $2$\,keV. At $2$\,keV energy, the AR values overestimate the measurements of \cite{Dagnac1970} by a factor $3.8$ on average; we chose to use the scaled AR cross section to estimate the low-energy dependence below $1.5$\,keV.
        \item \textbf{Fit and validity.} A polynomial fit of order $4$ in $\log_{10}$ was performed on the chosen (H, H$_2$O) electron-loss cross sections. The expected validity range is $\varv_i = 150-2\times 10^4$\,\kms{} ($0.1-2100$\,keV). This simple fit compares well to that performed by \cite{Uehara2000}. Confidence: medium. 
        \item \textbf{Further work.} Need for measurements for $0.05<E_i<5$\,keV. 
    \end{itemize}
    
        \item Reaction $\sigma_{0-1}$ ($\textnormal{H}^{0}\rightarrow\textnormal{H}^{-}$) 
        \begin{itemize}
            \item \textbf{Measurements.} State-selective time-of-flight measurements of the one-electron capture cross section for H in water were recently made by \cite{Luna2007} in the $8-100$\,keV range, which likely is above the cross-section peak. 
        \item \textbf{Uncertainty.} Uncertainties are on average $10$\%. 
        \item \textbf{Selection.} Between $8$ and $100$\,keV, we adopted the summed cross section over all target product channels of \cite{Luna2007}. To extend these measurements, we chose to use the additive rule for H$_2$ and O, that is, at low energies, data from \cite{Gealy1987_H} for H on H$_2$ paired with data from \cite{VanZyl2014} for H on O. At high energies, we used the measurements of \cite{Hill1979} in H$_2$ and those of \cite{Williams1984} in O. Finally, we scaled the overall reconstructed H$_2$+O data points to reach the magnitude of the \cite{Luna2007} data using a varying multiplication factor $1.3-4.8$ that depends on energy between $8$ and $30$\,keV, and a constant $\times4.8$ factor below $8$\,keV. 
        \item \textbf{Fit and validity.} A polynomial fit of order $5$ on the reconstructed dataset. Validity range $\varv_i = 100-4500$ \kms{} ($0.05-105$\,keV). Confidence: medium (low below $8$\,keV, high above). 
        \item \textbf{Further work.} Need for measurements at energies below the peak, for $0.05<E_i<10$\,keV.
    \end{itemize}
\end{itemize}

\subsubsection{H$^-$ -- H$_2$O}
The reactions involving the negative fast ion H$^-$ are the two-electron $\sigma_{-11}$ and the one-electron $\sigma_{-10}$ losses. They are shown in Fig.~\ref{fig:crossSection_CX_H} (right).
\begin{itemize}[label=$\bullet$]
        \item Reaction $\sigma_{-11}$ ($\textnormal{H}^{-}\rightarrow\textnormal{H}^{+}$) 
        \begin{itemize}
        \item \textbf{Measurements.} The only measurement found for the two-electron loss by H$^-$ in H$_2$O is that of \cite{Baribaud1971}, who reported a single cross section at $5$\,keV for H$_2$O, $\sigma_{-11}=0.7\times 10^{-20}$\,m$^2$. 
        \item \textbf{Uncertainty.} The reported error is about $30$\% at $5$\,keV. 
        \item \textbf{Selection.} Because of the lack of data, we adopted the additive rule $\sigma_{-11}$(H$_2$)+$\sigma_{-11}$(O$_2$)/2. For H$_2$, we used data from \cite{Geddes1980} in the energy range $1-300$\,keV \citep[dataset in excellent agreement for $\sigma_{-10}$ with that of][thus giving good confidence on their $\sigma_{-11}$ values]{Gealy1987_H}. For O$_2$, we used data reported by \cite{Williams1984} for $2.5<E_i<5$\,keV, \cite{Fogel1957} and by \cite{Lichtenberg1980} for $E_i=50-227$\,keV, which agree well in their common ranges. 
        \item \textbf{Fit and validity.} A polynomial fit of degree $4$ in $\log_{10}$ on the reconstructed (H$^-$,~H$_2$O) two-electron-loss cross section. At $5$\,keV, the additive rule fit is within $5\%$ of the reported value for H$_2$O \citep{Baribaud1971}. Validity range $\varv_i = 650-7600$\,\kms{} ($2.2-300$\,keV). Confidence: low.  
        \item \textbf{Further work.} Need for measurements at any energy, in priority in the energy range $0.1-100$\,keV.
    \end{itemize}
    
    \item Reaction $\sigma_{-10}$ ($\textnormal{H}^{-}\rightarrow\textnormal{H}^{0}$) 
        \begin{itemize}
        \item \textbf{Measurements.} The only measurement found for the one-electron loss by H$^-$ in H$_2$O is that of \cite{Baribaud1971} \citep[also in][]{Baribaud1972}, who reported a unique value at $5$\,keV in H$_2$O, $\sigma_{-10}=7.5\times 10^{-20}$\,m$^2$. 
        \item \textbf{Uncertainty.} The reported error is $13$\% at $5$\,keV. 
        \item \textbf{Selection.} Because of the lack of data, we chose to use the additive rule, $\sigma_{-10}$(H$_2$)+$\sigma_{-10}$(O$_2$)/2 and scaled it to the value of \cite{Baribaud1971} at $5$\,keV. For H$_2$, the data from \cite{Geddes1980} ($1-300$\,keV) and \cite{Hvelplund1982elecloss} ($300-3500$\,keV) were joined. For O, data from \cite{Williams1984} ($2.5-250$\,keV), which compare well with those from \cite{Lichtenberg1980} ($50-225$\,keV), and \cite{Rose1958} ($400-1500$\,keV) were adopted. At very low collision velocity, the energy of the center of mass is different from that of the ion energy measured in the laboratory frame. \cite{Huq1983elecdetach} and \cite{Risley1974}, reported by \cite{Phelps1990}, measured H$^-$ total electron loss in H$_2$ from a threshold at $2.38$\,eV to $200$\,eV ($\varv_i = 21-195$\,\kms{}), and from $300$\,eV to $10$\,keV ($240-1400$\,\kms{}), respectively. We note that in this energy range, single charge transfer dominates so that neutral hydrogen and negative molecular hydrogen ions are simultaneously produced: H$^-$+H$_2\rightarrow$H+H$_2^-$ \citep{Huq1983elecdetach}. Correspondingly, \cite{Bailey1970} made measurements in O$_2$ in the range $0.007-0.34$\,keV ($36-81$\,\kms{}), with values of about $10^{-19}$\,m$^2$/atom.
        Compared to the one reported value for H$_2$O at $5$\,keV, the reconstructed additive rule cross section overestimates the efficiency of the electron detachment by a factor $2.3$, which we chose as our scaling factor. The validity of such a scaling at one energy to extrapolate the values at other energies is likely subject to large uncertainties, which cannot be precisely assessed for lack of experimental or theoretical data.
        \item \textbf{Fit and validity.} A polynomial fit of degree $5$ in $\log_{10}$ on the reconstructed (H$^-$,~H$_2$O) one-electron-loss cross section, scaled to the value of \cite{Baribaud1971} at $5$\,keV. Validity range $\varv_i = 100-20\,000$\,\kms{} ($E_i = 0.05-2100$\,keV). Confidence: low. 
        \item \textbf{Further work.} Need for measurements at any energy, in priority above threshold, so that $0.1<E_i<100$\,keV.
    \end{itemize}
\end{itemize}

\subsubsection{Discussion}
Figure~\ref{fig:crossSection_CX_H} shows the charge-changing cross sections for (H$^+$,~H,~H$^-$). The dominant process below about $2000$\,\kms{} solar wind speed is electron capture $\sigma_{10}$ of H$^+$, which reaches a maximum value of about $2\times10^{-19}$\,m$^2$. A second process of importance is electron stripping $\sigma_{-10}$ of H$^-$ , reaching $0.8\times10^{-19}$\,m$^2$ at its peak at $400$\,\kms{}. However, since only one measurement has been reported in water for this process, the additive rule is likely to give only a crude approximation at low speeds; that said, because H$^-$ anions are populated by two very inefficient processes, this will likely result in a very small overall effect in the charge-state distributions (see Paper~II).
Consequently, at typical solar wind speeds, single-electron captures by H$^+$ and H are expected to drive the solar wind charge- state distribution in a water gas.

\section{Experimental ionization cross sections for (H,~He) in H$_2$O}\label{sec:IonisationXsections}

We present in this section the total ionization cross sections for the collisions of helium and hydrogen species with water molecules. 
Reviews at very high energies have been published over the past two decades with the development of Monte Carlo track-structure models describing how radiation interacts with biological tissues \citep{Uehara2002,Nikjoo2012}.

Ionization cross sections are noted $\sigma_{ii}$ , where the initial charge state $i$ stays the same during the reaction (target ionization only). The reactions we consider in this section are thus

\begin{align}
    \sigma_{ii}:&\ \textnormal{X}^{i+} + \textnormal{M} \longrightarrow \textnormal{X}^{i+} + \textnormal{[M]}^{q+} + qe^- \quad(\textnormal{X}^{i+}\rightarrow\textnormal{X}^{i+}),
\end{align}

with $q$ the number of electrons ejected from the neutral molecule M by a fast-incoming particle X. Because the initial solar wind ion distribution becomes fractionated on its path toward the inner cometary regions as a result of charge-transfer reactions, helium and hydrogen species are usually found in three charge states, namely X$^{i+}$, X$^{(i-1)+}$ and X$^{(i-2)+}$, with $i$ the charge of the species. Because of the detection methods we used, experimentally reported cross sections are usually total electron production cross sections or positive-ion production cross sections \citep{Rudd1985_Hepp,Gobet2006,Luna2007}, which may contain contamination from transfer-ionization processes (as the overall charge is conserved). For protons in water, these charge-transfer processes are, for example,

\begin{align*}
    \textnormal{H}^{+} + \textnormal{H}_2\textnormal{O} \longrightarrow \textnormal{H} + [\textnormal{H}_2\textnormal{O}]^{2+} + e^-.
\end{align*}

The contribution of charge-transfer processes to the measured cross section may become non-negligible at low energies. However, at typical solar wind energies, the total electron production cross sections decrease rapidly as power laws, making the transfer-ionization contribution small in comparison to any of the single or double charge-changing reactions considered in Sect.~\ref{sec:CXXsections}. When  the total charge-exchange and ionization rates are calculated from these two sets of cross sections, counting these minor charge-exchange reactions twice (a first time in the charge-exchange cross section and second time in the ionization) will therefore be minimized.

In ionization processes, the molecular target species M can also be dissociated into ionized fragments: for H$_2$O targets, ionization may lead to the formation of singly charged ions H$^+$, H$_2^+$, O$^+$ and OH$^+$ or even to that of doubly charged ions \citep[e.g., O$^{2+}$, as in][]{Werner1995}. 
In this section we only consider the total ionization cross section, which includes all dissociation paths of the target species, noted [M]$^{q+}$.

\subsection{Helium + H$_2$O}
\begin{figure*}
  \includegraphics[width=\linewidth]{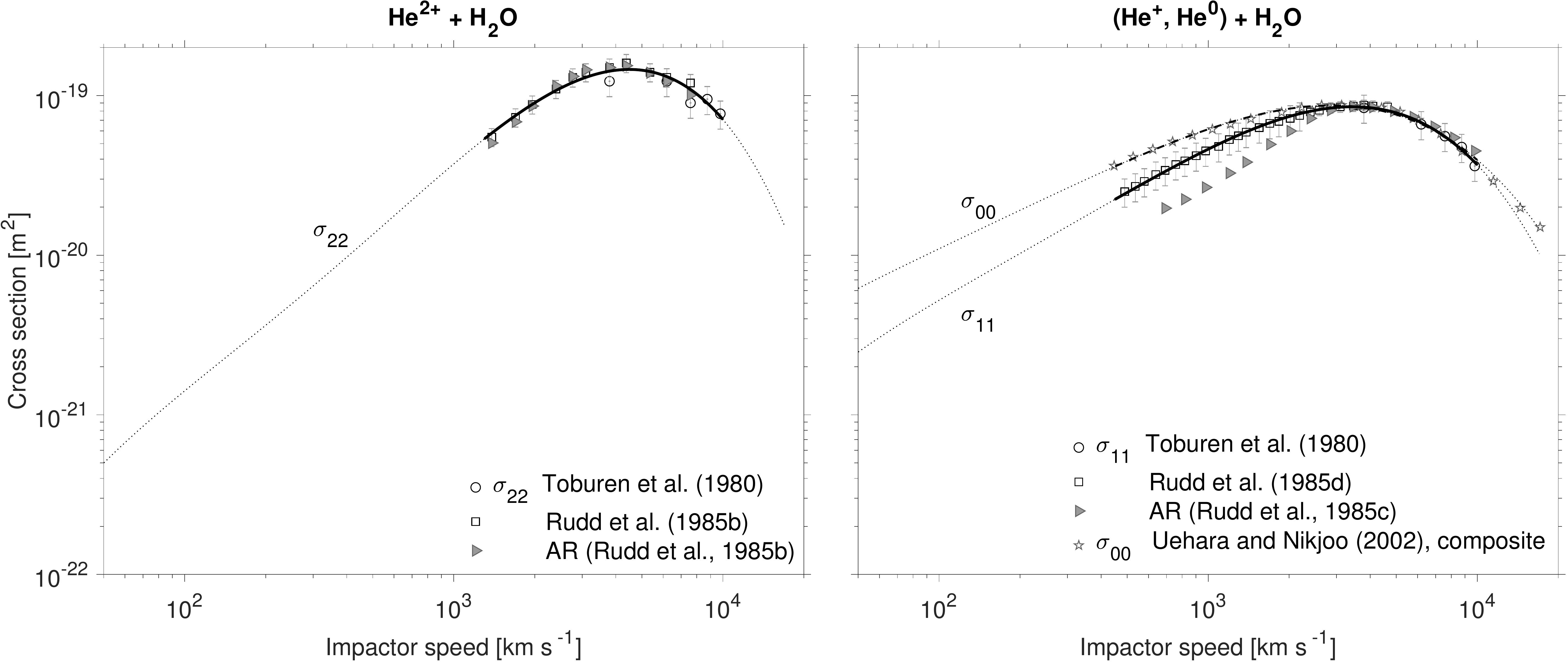}
     \caption{Experimental ionization cross sections for fast helium atoms and ions in a water gas as a function of impact speed. "AR" refers to the additive rule: when no experimental results for H$_2$O are available, results for H$_2$ and O$_2$ are combined to give an estimate (see text for details); experimental uncertainties for these estimates are at least $25$\%. As no experimental data are available for He$^0$, the composite recommendation of \cite{Uehara2002} was chosen (see text). Polynomial fits in thick continuous and dashed lines are also shown, and the coefficients are listed in Table~\ref{tab:heliumIonXsections}. Smooth extrapolations in power laws at low and high energies are indicated as thin dotted lines.
     }
     \label{fig:crossSection_Ion_He}
 \end{figure*}

Energies are given in keV per atomic mass unit (keV/u). Ionization cross sections for helium species are shown in Fig.~\ref{fig:crossSection_Ion_He}. Corresponding polynomial fit parameters are given in Table~\ref{tab:heliumIonXsections}.
\begin{itemize}[label=$\bullet$]
    \item Reaction $\sigma_{22}$ ($\textnormal{He}^{2+}\rightarrow\textnormal{He}^{2+}$) 
        \begin{itemize}
        \item \textbf{Measurements.} Laboratory measurements were performed by \cite{Rudd1985_Hepp} between $10$ and $300$\,keV/u ($1400-7500$\,\kms{}) for the total electron production, and by \cite{Toburen1980} between $75$\,keV/u and $500$\,keV/u.
        The additive rule $\sigma$(H$_2)+\sigma$(O) with the datasets of \cite{Rudd1985_Hepp} in the same energy range yields results in excellent agreement with the water measurements (no measurements in H$_2$ and O$_2$ below $10$\,keV were found).
        \item \textbf{Uncertainties.} Uncertainties are $13\%$ below $300$\,keV/u \citep{Rudd1985_Hepp}, and reach $20\%$ above \citep{Toburen1980}.
        \item \textbf{Selection.} Since the datasets are complementary in energy and agree well with each other, we used both water measurements. 
        \item \textbf{Fit and validity.} A polynomial fit of order $4$ of the cross section as a function of the logarithm of the impact speed was performed. Expected validity range is $\varv_i = 1400-10\,000$\,\kms{} ($E_i = 10-520$\,keV/u). Confidence: high.
        \item \textbf{Further work.} Need for low-energy ($E_i<10$\,keV/u) and very high-energy ($E_i>500$\,keV/u) measurements.
    \end{itemize}
    
    \item Reaction $\sigma_{11}$ ($\textnormal{He}^{+}\rightarrow\textnormal{He}^{+}$) 
        \begin{itemize}
            \item  \textbf{Measurements.} Ionization cross sections have been measured by \cite{Rudd1985_Hep} between $1.25$ and $112.5$\,keV/u ($490-4650$\,\kms{}), and by \cite{Toburen1980} between $75$\,keV/u and $500$\,keV/u. 
            The additive rule using the measurements of \cite{Rudd1985_Hep_gases} in H$_2$ and O$_2$ agrees well at the cross-section peak and above ($>1$\,keV/u) but increasingly diverges below (up to a factor $2$).
        \item \textbf{Uncertainties.} Uncertainties are $20\%$ below $30$\,keV/u and lower than $8\%$ in the $30-450$\,keV/u range \citep{Rudd1985_Hep}. At energies above $450$\,keV/u, errors are on the order of $20\%$ \citep{Toburen1980}.
        \item \textbf{Selection.} The two H$_2$O datasets overlap with each other and are in excellent agreement. We therefore used both datasets.
        \item \textbf{Fit and validity.} A polynomial fit of order $4$ in $\log_{10} \varv_i$ was performed. Expected validity is $450-10\,000$\,\kms{} ($E_i = 1-520$\,keV/u). Confidence: high.
        \item \textbf{Further work.} Need for very low-energy ($E_i<1$\,keV/u) and very high-energy ($E_i>500$\,keV/u) measurements.
    \end{itemize}
    
    \item Reaction $\sigma_{00}$ ($\textnormal{He}^{0}\rightarrow\textnormal{He}^{0}$) 
        \begin{itemize}
        \item \textbf{Measurements.} No measurement of He$^0$ impact ionization on water has been performed.   
        \item \textbf{Uncertainties.} -
        \item \textbf{Selection.} To palliate the lack of measurements, \cite{Uehara2002} \citep[reported in][with no alterations]{Nikjoo2012} proceeded in two steps with their Monte Carlo track-structure numerical model: at low energies (below $100$\,keV/u), where He$^0$ atoms dominate the composition of the charge distribution as a result of charge exchange, the authors adjusted the total ionization cross sections of He$^0$+H$_2$O to match the total electronic stopping powers of the helium system tabulated in report $49$ of ICRU \citep{ICRU1993}. At energies above $100$\,keV/u, ionization cross sections of He$^0$ were assumed to be equal to those of He$^+$ measured by \cite{Toburen1980}. Expected uncertainties according to \cite{Uehara2002} are of the order of $20\%$. These cross sections were chosen here. However, because no specific measurements have been made, we ascribe a low confidence level to this estimate, especially at typical solar wind energies. 
        \item \textbf{Fit and validity.} A polynomial fit in $\log_{10} \varv_i$ was performed. Expected validity range for such a composite estimate is $100-10\,000$\,\kms{} ($E_i = 0.05-520$\,keV/u). Confidence: low.
        \item \textbf{Further work.} Measurements at any energy ($E_i>0.05$\,keV/u) is needed.
    \end{itemize}
\end{itemize}

 \begin{table*}
        \centering
        \caption{Recommended ionization cross-section polynomial fits for (He$^{2+}$,~He$^+$,~He$^0$) projectiles colliding with H$_2$O vapor. }
        \label{tab:heliumIonXsections}
    \small
        \begin{tabular}{lcrrrrrrllc} 
                \hline\hline
        Cross section & Degree &  \multicolumn{5}{c}{Coefficients}  & \multicolumn{2}{c}{Validity range} & Confidence\\ \relax
                [m$^{2}$] & $n$ & $p_0$ & $p_1$ & $p_2$ & $p_3$ & $p_4$ & $\varv_i$ [\kms{}] & $E_i$ [keV/u] & \\
                \hline
        $\sigma_{22}$ & $4$ & $-265.3697$ & $176.7016$ & $-48.3734$ & $5.91427$ & $-0.27030$ & $1400-10\,000$ & $10.0-520$ & high \\  
        $\sigma_{11}$ & $4$ & $-169.8607$ & $108.1412$ & $-29.7492$ & $3.66917$ & $-0.16967$ & $450-10\,000$ & $1.06-520$ & high \\  
        $\sigma_{00}$ & $4$ & $-87.3445$ & $49.6893$ & $-14.1573$ & $1.82417$ & $-0.08824$ & $450-10\,000$ & $1.06-520$ & low \\
                \hline
        \end{tabular}
        \tablefoot{The polynomial, function of the speed of the impactor, is of the form $\log_{10}(\sigma) = \sum_{j=0}^n{p_j\,(\log_{10} \varv_i)^{j}}$, where $n$ is the degree of the fit, the speed $\varv_i$ is expressed in m\,s$^{-1}$ and the cross section $\sigma$ in m$^2$. Ranges of validity for impact speeds and energies are given. Confidence levels on the fits are indicated as high ($<25\%$), medium ($25-75\%$), and low ($>75\%$) (see text).}
 \end{table*}

\subsection{Hydrogen + H$_2$O}
\begin{figure*}
  \includegraphics[width=\linewidth]{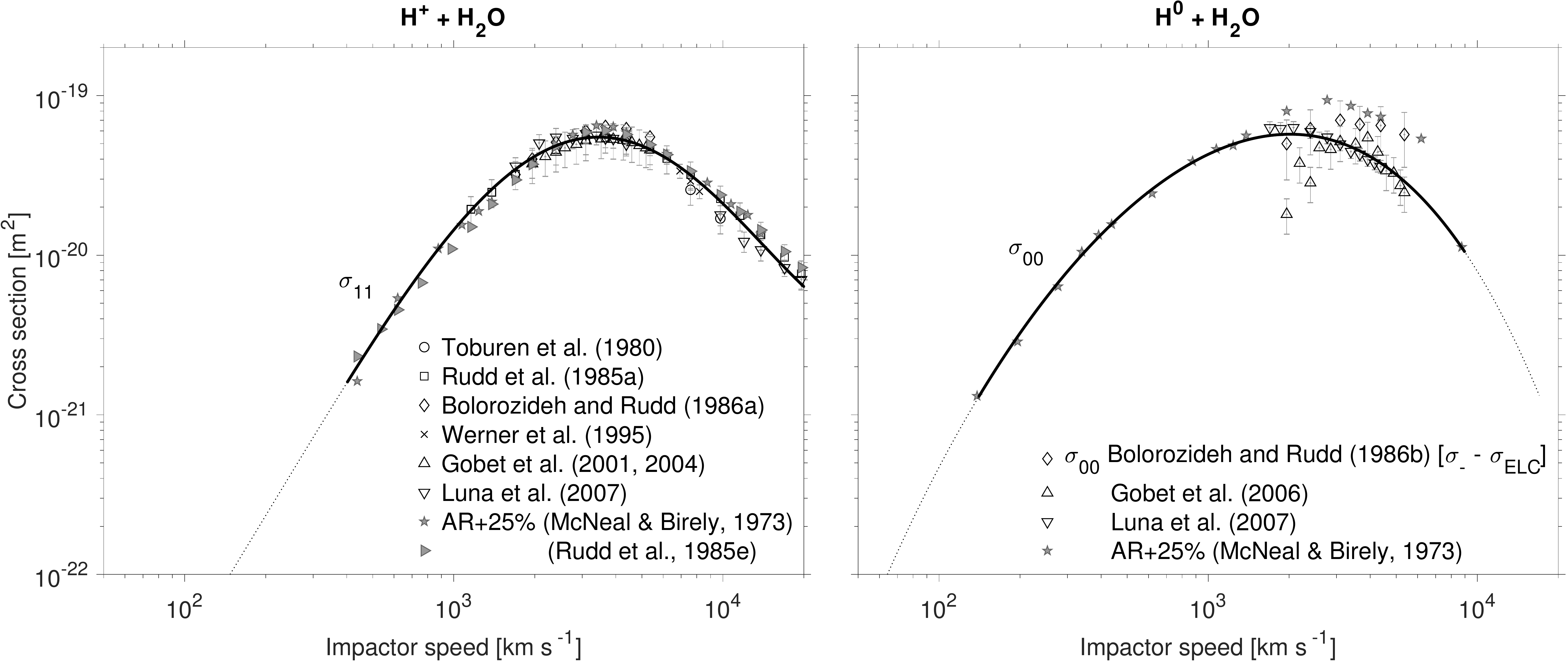}
     \caption{Experimental ionization cross sections for fast hydrogen atoms and protons in a water gas as a function of impact speed. "AR" refers to the additive rule: when no experimental results for H$_2$O are available, results for H$_2$ and O$_2$ are combined to give an estimate (see text for details); experimental uncertainties for these estimates are at least $25$\%. No measurement for the ionization of H$_2$O by H$^-$ is available. Polynomial fits in thick lines are also shown, and the coefficients are listed in Table~\ref{tab:hydrogenIonXsections}. Smooth extrapolations in power laws at low and high energies are indicated as thin dotted lines.
     }
     \label{fig:crossSection_Ion_H}
 \end{figure*}

Ionization cross sections for hydrogen species are shown in Fig.~\ref{fig:crossSection_Ion_H}. Corresponding polynomial fit parameters are given in Table~\ref{tab:hydrogenIonXsections}.
\begin{itemize}[label=$\bullet$]
        \item Reaction $\sigma_{11}$ ($\textnormal{H}^{+}\rightarrow\textnormal{H}^{+}$) 
        \begin{itemize}
            \item $\sigma_{11}$. \textbf{Measurements.} Over the past three decades, many experiments have been carried out on the ionization of water by fast protons. \cite{Toburen1980} compared their results at high energies for helium particles to those of protons \citep{Toburen1977} at $300$ and $500$\,keV; proton cross sections were found to be about half those of He$^+$. \cite{Rudd1985_Hp_H2O} measured cross sections in the range $7-4000$\,keV and \cite{Bolorizadeh1986_Hp} in the $15-150$\,keV range. These datasets are in good agreement with the more recent total ionization measurements by \cite{Werner1995} between $100$ and $400$\,keV and those of \cite{Gobet2001} and \cite{Gobet2004}, who focused on the production of dissociation fragments at $20$ to $400$\,keV energies. \cite{Luna2007} reported total and partial ionization cross sections between $15-100$ and $500-3500$\,keV; in their lower energy range, and similar to \cite{Werner1995}, their total ionization cross sections also include contributions from the transfer ionization reaction $\textnormal{H}^+ +\textnormal{H}_2\textnormal{O}\rightarrow \textnormal{H} + \textnormal{H}_2\textnormal{O}^{2+} + e$. When this is compared to the direct ionization measurements of \cite{Gobet2001} and \cite{Gobet2004}, it appears that the total ionization cross section is probably not strongly affected by this additional contribution.
        \item \textbf{Uncertainties.} Uncertainties are about $20\%$ between $7-15$\,keV energy \citep{Rudd1985_Hp_H2O}, and lower than $15\%$ at energies above $15$\,keV \citep{Luna2007}. 
        \item \textbf{Selection.} We performed fits on all measurements listed above between $7$ and $4000$\,keV. To extend the dataset to lower energies, the additive rule with the measurements of \cite{McNeal1973} between $2-800$\,keV and \cite{Rudd1985_Hp_gases} between $1-5000$\,keV  was used and scaled to match the results of \cite{Rudd1985_Hp_H2O} at the cross-section peak; the composite $\sigma$(H$_2$)+$\sigma$(O$_2)/2$ cross section is on average $25\%$ smaller than the direct H$_2$O measurements.
        \item \textbf{Fit and validity.} Polynomial fit of degree $5$ was performed on the reconstructed dataset. Expected validity range is $400-10\,000$\,\kms{} ($E_i = 0.8-520$\,keV/u). Confidence: high.
        \item \textbf{Further work.} Need of laboratory measurements at low energies to very low energies ($0.05 < E_i < 10$\,keV).
    \end{itemize}

    \item Reaction $\sigma_{00}$ ($\textnormal{H}^{0}\rightarrow\textnormal{H}^{0}$) 
        \begin{itemize}
        \item \textbf{Measurements.} Laboratory measurements for the ionization of H$_2$O by fast hydrogen energetic neutral atoms (ENAs) were performed by \cite{Bolorizadeh1986_H} between $20$ and $150$\,keV. Electron loss to the continuum (ELC) cross sections were also concurrently calculated, which need to be subtracted from the total electron production cross sections (marked "$\sigma_{-}$" in the terminology of Rudd's team), as explained in detail in \cite{Gobet2006}. Recently, the total target ionization cross section was measured by \cite{Gobet2006} using time-of-flight coincidence and imaging techniques and by \cite{Luna2007} using time-of-flight mass analysis, both above $15-20$\,keV impact energy. As pointed out by \cite{Luna2007}, the measurements of \cite{Gobet2006} at low collision energies are likely to have missed a portion of the proton beam scattered at high angles, suggesting an underestimation of their signal below about $30$\,keV. Consequently, the measurements of \cite{Gobet2006} and \cite{Luna2007} diverge by a factor $2-3$ below $30$\,keV, but they agree well above this limit. The ELC-corrected measurements of \cite{Bolorizadeh1986_H} are larger by a factor $1.4-2$ above $50$\,keV.   
        \item \textbf{Uncertainties.} Uncertainties are quoted to be about $20\%$ below $15$\,keV by \cite{Bolorizadeh1986_H}, whereas \cite{Luna2007} claimed errors of about $10\%$ at all energies probed. Following \cite{Luna2007}, we give the measurements by \cite{Gobet2006} a high uncertainty of $25\%$ because of the uncertainty in their calibration.
        \item \textbf{Selection.} We chose the datasets of \cite{Luna2007} ($15-150$\,keV) and that of \cite{Gobet2006}, which is restricted to $50-100$\,keV energies. To approximate the low-energy and high-energy dependence of the cross section, we used the additive rule $\sigma$(H$_2$)+$\sigma$(O$_2)/2$ with the datasets of \cite{McNeal1973} between $0.1$ and $10$\,keV, upscaled by $25\%$, as in the case of proton ionization cross sections (see $\sigma_{11}$ case above). This results in a smooth decrease at low energy, a trend that cannot be extrapolated further, however. 
        \item \textbf{Fit and validity.}  A polynomial fit of degree $4$ was performed on the reconstructed dataset. Expected validity range: $140-9000$\,\kms{} ($E_i = 0.1-420$\,keV/u). Confidence: low (low at $\varv<1000$\,\kms{}, medium above).
        \item \textbf{Further work.} Need of laboratory measurements at any energy in the range $0.05-500$\,keV.  
    \end{itemize}

    \item Reaction $\sigma_{-1-1}$ ($\textnormal{H}^{-}\rightarrow\textnormal{H}^{-}$) 
        \begin{itemize}
            \item \textbf{Measurements}. No measurement of impact ionization cross sections in collisions of H$^-$ with H$_2$O has been performed or studied, to our knowledge. Similarly, no measurement has been found of ionization of H and O, as separate entities. Because charge fractions at comets do not favor the presence of H$^-$ and because of the low expected fluxes, we avoid speculation. This species and its associated ionization cross section in water is left for further laboratory studies.
    \end{itemize}
\end{itemize}

 \begin{table*}
        \centering
        \caption{Recommended ionization cross-section polynomial fits for (H$^{+}$,~H$^0$) projectiles colliding with H$_2$O vapor.}
        \label{tab:hydrogenIonXsections}
    \tiny
        \begin{tabular}{lcrrrrrrllc} 
                \hline\hline
        Cross section & Degree &  \multicolumn{6}{c}{Coefficients}  & \multicolumn{2}{c}{Validity range} & Confidence\\ \relax
                [m$^{2}$] & $n$ & $p_0$ & $p_1$ & $p_2$ & $p_3$ & $p_4$ & $p_5$ & $\varv_i$ [\kms{}] & $E_i$ [keV/u] & \\
                \hline
                $\sigma_{11}$   & $5$ & $-1437.4092$ & $1233.0652$ & $-429.3680$ & $74.35337$ & $-6.37862$ & $0.216408$ & $400-30\,000$ & $0.84-4700$ & high \\
                $\sigma_{00}$   & $4$ & $-205.9851$ & $114.3405$ & $-27.6772$ & $3.13079$ & $-0.13835$ & - & $140-10\,000$ & $0.10-420$  
& low \\
                $\sigma_{-1-1}$ & - & - & - & - & - & - & - & - & - &  \emph{no data} \\
                \hline
        \end{tabular}
        \tablefoot{The polynomial, function of the speed of the impactor, is of the form $\log_{10}(\sigma) = \sum_{j=0}^n{p_j\,(\log_{10} \varv_i)^{j}}$, where $n$ is the degree of the fit, the speed $\varv_i$ is expressed in m\,s$^{-1}$ and the cross section $\sigma$ in m$^2$. Ranges of validity for impact speeds and energies are given. Confidence levels on the fits are indicated as high ($<25\%$), medium ($25-75\%$), and low ($>75\%$) (see text).}
 \end{table*}

\subsection{Discussion}
Figures~\ref{fig:crossSection_Ion_He} and \ref{fig:crossSection_Ion_H} show the total ionization cross sections for (He$^{2+}$,~He$^+$,~He$^0$) and (H$^{+}$,~H$^0$). The ionization cross sections for helium species are on average $1.5-2.5$ times larger than those for the hydrogen system. They peak for helium at $1.5\times10^{-19}$\,m$^2$ around $5000$\,\kms{} ($0.8\times10^{-19}$\,m$^2$, $2850$\,\kms{}) for $\sigma_{22}$ ($\sigma_{11}$, respectively). For hydrogen, the cross sections peak at $0.5\times10^{-19}$\,m$^2$ around $3450$\,\kms{} ($0.6\times10^{-19}$\,m$^2$, $1950$\,\kms{}) for $\sigma_{11}$ ($\sigma_{00}$). Consequently, the largest cross sections are encountered for the higher charge states of each system of species. However, for all ionization cross sections, their low-energy dependence is subject to large uncertainties as a result of a lack of experimental results.

\section{Recommended Maxwellian-averaged cross sections for 
solar wind-cometary interactions}\label{sec:maxwell}

In the interplanetary medium, solar wind ion velocity distributions may be approximated by a Maxwellian at a constant temperature $T_p$ that typically is around $10^5$\,K \citep{Meyer2012}, or by a Kappa distribution in order to better take into account the tail of the velocity distribution \citep{Livadiotis2018}. For a Maxwellian distribution, which is a good first approximation, the temperature dependence with respect to heliocentric distance is such that $T_p = 8\times10^4\ R_\textnormal{Sun}^{-2/3}$ \citep{Slavin1981}, with $R_\textnormal{Sun}$ the heliocentric distance in AU. This corresponds to thermal speeds $\varv_\textnormal{th} = \sqrt{3\,k_BT_p/m_p}$ of about $45$\,\kms{} at $1$\,AU for protons of mass $m_p$, and half of that value for $\alpha$ particles: these values are much lower than the average "undisturbed" solar wind speed of $400$\,\kms{}. However, at the interface between the undisturbed solar wind and the cometary plasma environment, for instance at bow shock-like structures, considerable heating of the ions and electrons may occur simultaneously to the slowing down of the solar wind flow \citep{Koenders2013,CSW2017}. 
Solar wind proton and electron temperatures during planetary shocks can reach values up to several $10^6$\,K, for instance when interplanetary coronal mass ejections impact the induced magnetosphere of Venus \citep[$T_p$ increasing up to $100$\,eV, see][]{Vech2015}.  
\cite{Gunell2018} reported the first indication of a bow shock structure appearing for weak cometary outgassing rates at comet 67P: these structures were associated with a thermal spread of protons and $\alpha$ particles of a few $100$\,\kms{} ($T_p \gtrsim 10^6$\,K).
Deceleration and heating of the solar wind may in turn result in a spatially extended increase of the efficiency of the charge exchange and ionization \citep{Bodewits2006}. Bodewits and collaborators convolved a 3D drifting Maxwellian distribution with their He$^{2+}-$H$_2$O electron capture cross sections and found that for solar wind velocities below $400$\,\kms{} and a temperature above $10^6$\,K ($\varv_\textnormal{th} \sim 80$\,\kms{}), the cross section could be increased by a factor $2-10$.

For low-activity comets such as comet 67P, \cite{Behar2017} calculated the velocity distributions of solar wind protons measured during the in-bound leg of the \emph{Rosetta} mission. It is clear from their Fig.~2 that the velocity distribution functions cannot be approximated by a Maxwellian, which assumes dynamic and thermal equilibrium through collisions (ions thermalized at one temperature). Moreover, as CX reactions involve both the distribution of neutrals and ions, cross sections should be averaged over two velocity distributions, one for the neutrals, the other for the ions, with a reduced mass for the collision \citep{Banks1973a}. Therefore, the following development only gives an indicator of global effects for increased solar wind temperatures at a comet assuming a Maxwellian velocity distribution, without taking into account the measured angular and energy distributions of the solar wind ions.

\subsection{Method}

In order to take the effects of a Maxwellian velocity distribution for the solar wind into account, energy-dependent cross sections $\sigma_i$ for solar wind impacting species $i$ can be Maxwellian-averaged over all thermal velocities $\varv$ following the descriptions of \cite{Banks1973a} and \cite{Bodewits2006}:
\begin{align}
    \langle \sigma_i \rangle = \frac{\langle\sigma_i\, \varv\rangle}{\langle\varv\rangle}\ \hat{=}\     \sigma_{i,\textnormal{MACS}},
    \label{eq:macs}
\end{align}
defining the Maxwellian-averaged cross section (MACS), $\sigma_{i,\textnormal{MACS}}$. The 3D drifting Maxwellian velocity distribution is defined as
\begin{align}
    f_\textnormal{M}(\vec{v}) = \left(\frac{m_i}{2\pi\,k_BT_i}\right)^{3/2}\ \exp\left[-\frac{m_i}{2k_BT_i}\left(\vec{v}-\vec{v}_d\right)^2\right], 
\end{align}
with $\vec{v}_d$ the drift velocity of the solar wind (directed along its streamlines so that $\vec{v}_d = \vec{U_\textnormal{sw}}$, with $\vec{U_\textnormal{sw}}$ the solar wind velocity), $m_i$ the mass of the solar wind species considered, and $T_i$ its corresponding temperature. The distribution can be expressed in terms of $(\varv,\theta)$, with $\theta$ the angle between the solar wind drift and thermal velocities, so that the term in the exponential becomes $\left(\vec{v}-\vec{v}_d\right)^2=\varv^2+\varv_d^2-2\varv\,\varv_d\cos\theta$. Because the undisturbed solar wind is in a first approximation axisymmetric around its direction of drift propagation, angles are integrated between $0$ and $\pi$.

Consequently, the Maxwellian-averaged reaction rate $\langle \sigma_{i}\,\varv \rangle$ is \citep{Bodewits2006}
\begin{align}
\langle\sigma_i\, \varv\rangle =  2\pi\int_{0}^\infty{\int_0^\pi \sigma_i(\varv)\ \varv^3\ f_\textnormal{M}(\varv,\theta)\ \drv\varv \sin\theta\drv\theta},
\label{eq:sigmav_avg}
\end{align}
whereas the average speed $\langle \varv \rangle$ is
\begin{align}
    \langle \varv \rangle =  2\pi\int_{0}^\infty{\int_0^\pi \varv^3\ f_\textnormal{M}(\varv,\theta)\ \drv\varv \sin\theta\drv\theta}.
    \label{eq:v_avg}
\end{align}

Combining equations~(\ref{eq:sigmav_avg}) and (\ref{eq:v_avg}), we can calculate the MACS. The double integrals are solved numerically using the fitted polynomial functions of Sections~\ref{sec:CXXsections} and \ref{sec:IonisationXsections} by summing small contiguous velocity intervals logarithmically spaced from $10$ to $3000$\,\kms{} ($\text{three}$ times the maximum thermal velocity considered). Because the shape at low velocities of the final MACS in turn depends on the velocity dependence of the original cross sections and their smooth extrapolation below $100$\,\kms{}, the MACS are conservatively calculated in the restricted $100-800$\,\kms{} range. Extrapolations down to $10$\,\kms{} for the integration were made using power laws connecting at $120$\,\kms{}.

\subsection{Maxwellian-averaged charge-exchange cross sections}

To illustrate the effect of the non-monochromaticity of the solar wind, we calculated the effect on the cross sections of increasingly higher temperatures ($T=10^5$\,K, $1.6\times10^6$\,K, $10\times10^6$\,K, and $40\times10^6$\,K) corresponding to thermal velocities of $25$\,\kms{} ($50$\,\kms{}), $100$\,\kms{} ($200$\,\kms{}), $250$\,\kms{} ($500$\,\kms{}), and $500$\,\kms{} ($1000$\,\kms{}) for He$^{2+}$ (H$^+$, respectively). 
The three lowest temperatures are reasonably encountered in cometary environments, especially when the solar wind is heated following the formation of a shock-like structure upstream of the cometary nucleus.

\begin{figure*}
  \includegraphics[width=\linewidth]{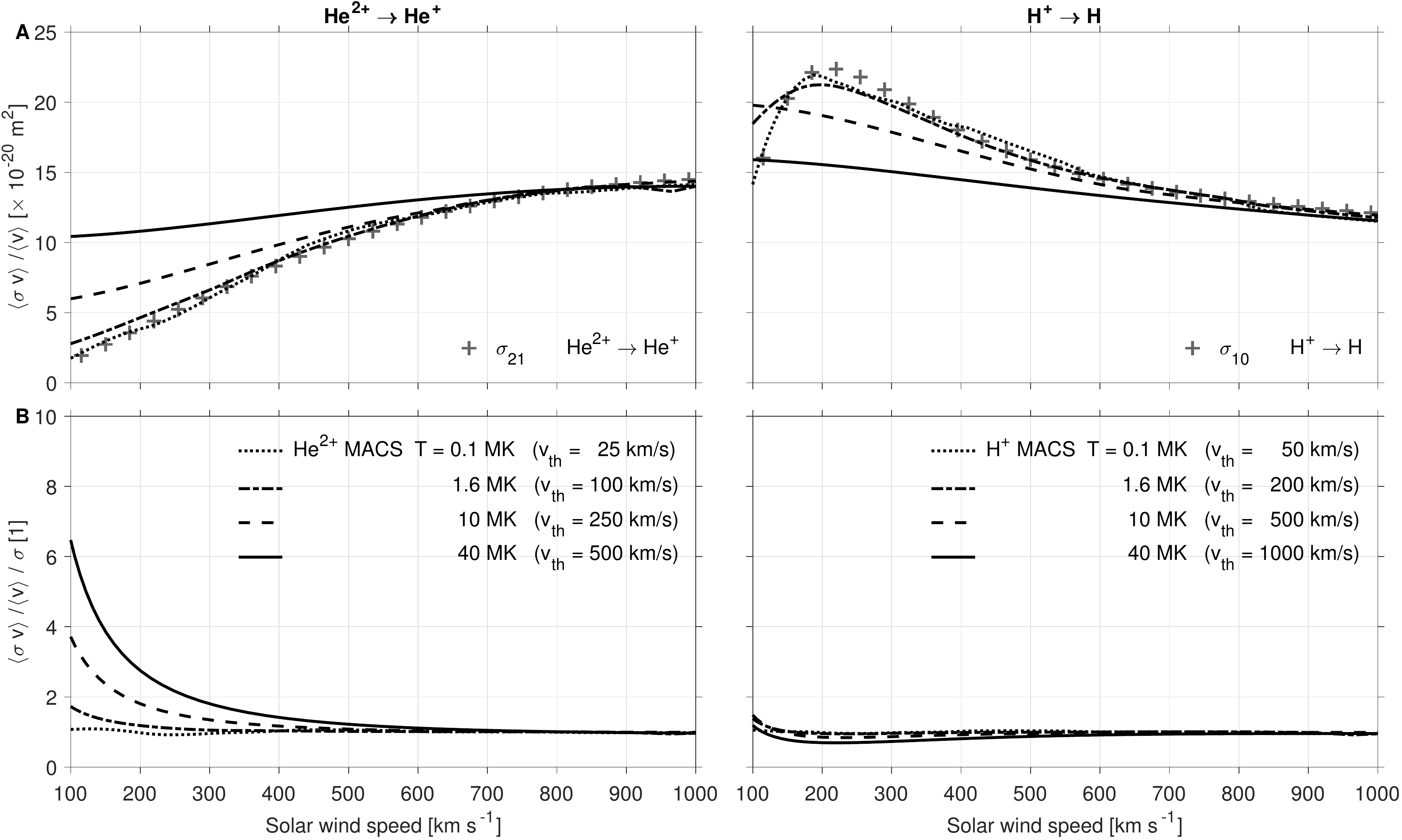}
     \caption{(A). Maxwellian-averaged cross section for single-electron capture by He$^{2+}$ (left) and H$^+$ (right) in H$_2$O. (B). Multiplication factor for the reference non-averaged cross sections. 
     Four solar wind temperatures were used: $T = 0.1$\,MK, $1.6$\,MK, $10$\,MK, and $40$\,MK, corresponding to varying thermal velocities.
     }
     \label{fig:crossSection_MACS_sigma21_sigma10}
 \end{figure*}

Figure~\ref{fig:crossSection_MACS_sigma21_sigma10}A shows the MACS for the two most important single-electron capture cross sections $\sigma_{21}$ (for $\alpha$ particles) and $\sigma_{10}$ (for protons). This illustrates that depending on which velocity the cross section peaks at, the Maxwellian-averaged cross section may be decreased or increased. This is easily understood because particles populating the tail of the Maxwellian at high velocities, where cross sections either peak or decrease in power law, will contribute to the averaged cross section. If the cross-section peak is located at high velocities, the MACS will be enhanced at low velocities (ratio of MACS to parent cross section higher than 1). Correspondingly, if the cross section peaks at low velocities, the MACS may become less than its initial cross section at low velocities (ratio lower than 1). This is demonstrated in Fig.~\ref{fig:crossSection_MACS_sigma21_sigma10}B, where the MACS is divided by its parent (non-averaged) cross section. In the case of hydrogen (Fig.~\ref{fig:crossSection_MACS_sigma21_sigma10}, right), the original $\sigma_{10}$ cross section peaks at $200$\,\kms{}: for any temperature of the solar wind, the MACS will thus be smaller than its parent cross section. Conversely, when the peak of the cross section is at higher velocities, as is the case for $\sigma_{21}$ (peak at $1000$\,\kms{}, Fig.~\ref{fig:crossSection_MACS_sigma21_sigma10}, left), the effect of the Maxwellian will be maximized and the MACS can become much larger than its parent cross section, depending on the temperature.

The result for helium is in qualitative agreement with that of \cite{Bodewits2006}: Maxwellian-induced effects start to become important at solar wind velocities below $400$\,\kms{} and for temperatures above $10^6$\,K.
For $500$\,\kms{} thermal velocities, the cross section is multiplied by a factor $6.5$ at a solar wind speed of $100$\,\kms{}. For lower thermal velocities ($250$, $100$ and $25$\,\kms{}), this multiplication factor decreases to $3.7$, $1.7,$ and $1.1$ at the same solar wind speed. Differences with the results of \cite{Bodewits2006} are due to different adopted cross sections.

For hydrogen, the cross section is almost unchanged until thermal velocities reach $1000$\,\kms{}, for which the MACS is minimum around $215$\,\kms{} solar wind speed (multiplication factor $0.7$ with respect to its parent cross section). A moderate increase is observed at a solar wind speed of $100$\,\kms{}, with factors ranging from $1.19$ ($\varv_\textnormal{th} = 1000$\,\kms{}), $1.50$ ($\varv_\textnormal{th} = 500$\,\kms{}), $1.38$ ($\varv_\textnormal{th} = 200$\,\kms{}), and $1.06$ ($\varv_\textnormal{th} = 50$\,\kms{}).

\subsection{Maxwellian-averaged ionization cross sections}
\begin{figure*}
  \includegraphics[width=\linewidth]{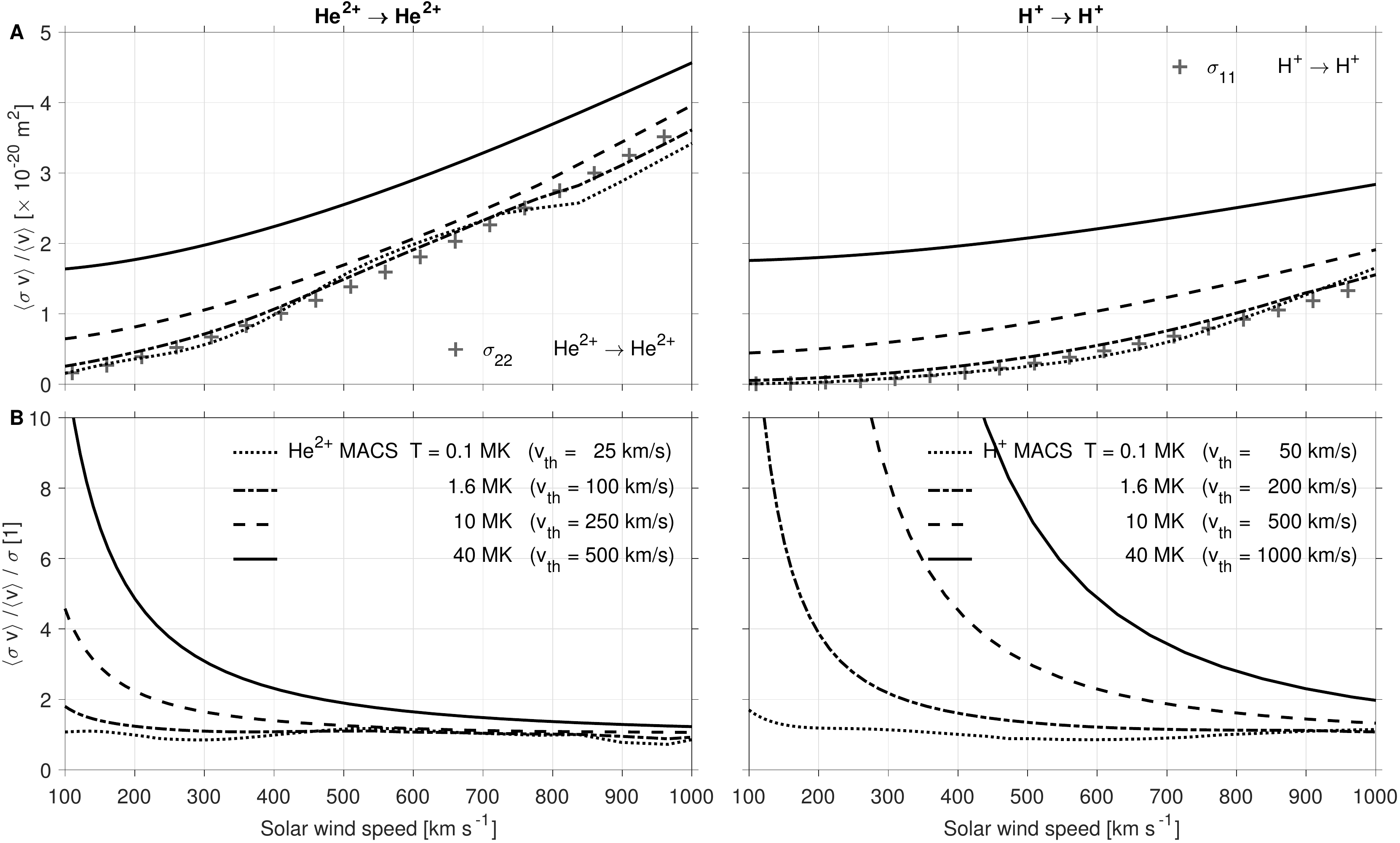}
     \caption{(A). Maxwellian-averaged cross section $\sigma_\textnormal{MACS}$ for total ionization of H$_2$O by impacting He$^{2+}$ (left) and H$^+$ (right). (B). Multiplication factor $\gamma$ for the reference non-averaged cross sections, i.e., $\gamma = \sigma_\textnormal{MACS} / \sigma$, for He$^{2+}$ (left) and H$^+$ (right). 
     Four solar wind temperatures were used: $T = 0.1$\,MK, $1.6$\,MK, $10$\,MK, and $40$\,MK, corresponding to varying thermal velocities.
    }
     \label{fig:crossSection_MACS_ionisation}
 \end{figure*}

We present here the MACS for ionization of water vapor by solar wind particles. Because ionization cross sections peak at speeds above $1000$\,\kms{}, the increase of the cross sections due to the tail of the Maxwellian distribution will be most important below $400$\,\kms{} for high solar wind temperatures. This is shown in Fig.~\ref{fig:crossSection_MACS_ionisation}, where Maxwellian-averaged ionization cross sections are calculated for He$^{2+}$ ($\sigma_{22}$) and H$^+$ ($\sigma_{11}$).

Cross sections are notably enhanced at typical solar wind speeds ($<1000$\,\kms{}), with an increase of more than a factor $10$ at a temperature $T = 40\times10^6$\,K ($\varv_\textnormal{th}=500$\,\kms{}) and a solar wind speed of $100$\,\kms{} for He$^{2+}$ (Fig.~\ref{fig:crossSection_MACS_ionisation}B, left). Correspondingly, the effect is even more drastic for H$^{+}$ (Fig.~\ref{fig:crossSection_MACS_ionisation}B, right), with a factor $10$ already reached at $400$\,\kms{} solar wind speed for a temperature of $40\times10^6$\,K, increasing up to a factor $500$ at $100$\,\kms{}. At a solar wind speed of $100$\,\kms{}, proton ionization cross sections are multiplied by a factor $16$ and $135$ at the more moderate temperatures of $1.6\times10^6$\,K and $10\times10^6$\,K, respectively.

\subsection{Recommended MACS}
Recommended Maxwellian-averaged cross-section bivariate polynomial fits were performed in the 2D solar wind temperature-speed ($T$,$\varv$) space. To constrain the fits, we restricted the temperature range to $0.1-6.4\times10^6$\,K for helium particles ($0.025-10\times10^6$\,K for hydrogen particles) and set a common solar wind velocity range of $100-800$\,\kms{}. We first used a moving average on the calculated MACS to smooth out the sometimes abrupt variations with respect to solar wind speed, which are due to the numerical integration.

Bivariate polynomials of degree $(n,m)$ in temperature (degree $n$) and velocity (degree $m$) can be written for the MACS as $\sigma_\textnormal{MACS}(T,\varv)~=~\sum_{i,j}{~p_{ij}~T^i~\varv^j}~\times~10^{-22}$ [m$^2$].
Least-absolute-residuals (LAR) bivariate polynomial fits of degree $(4,4)$ or $(4,5)$, with $15$ or $20$ coefficients, respectively, were performed so that the fits could apply to the maximum number of cases without losing in generality and fitting accuracy. The LAR method was preferred to least squares as it gives less weight to extreme values, which, despite smoothing, may appear and are usually connected to the velocity discretization used for the MACS integration. Appendix~\ref{appendix:MACSfits} presents our recommended bivariate least-squares polynomial fitting coefficients.

\begin{figure*}
  \includegraphics[width=\linewidth]{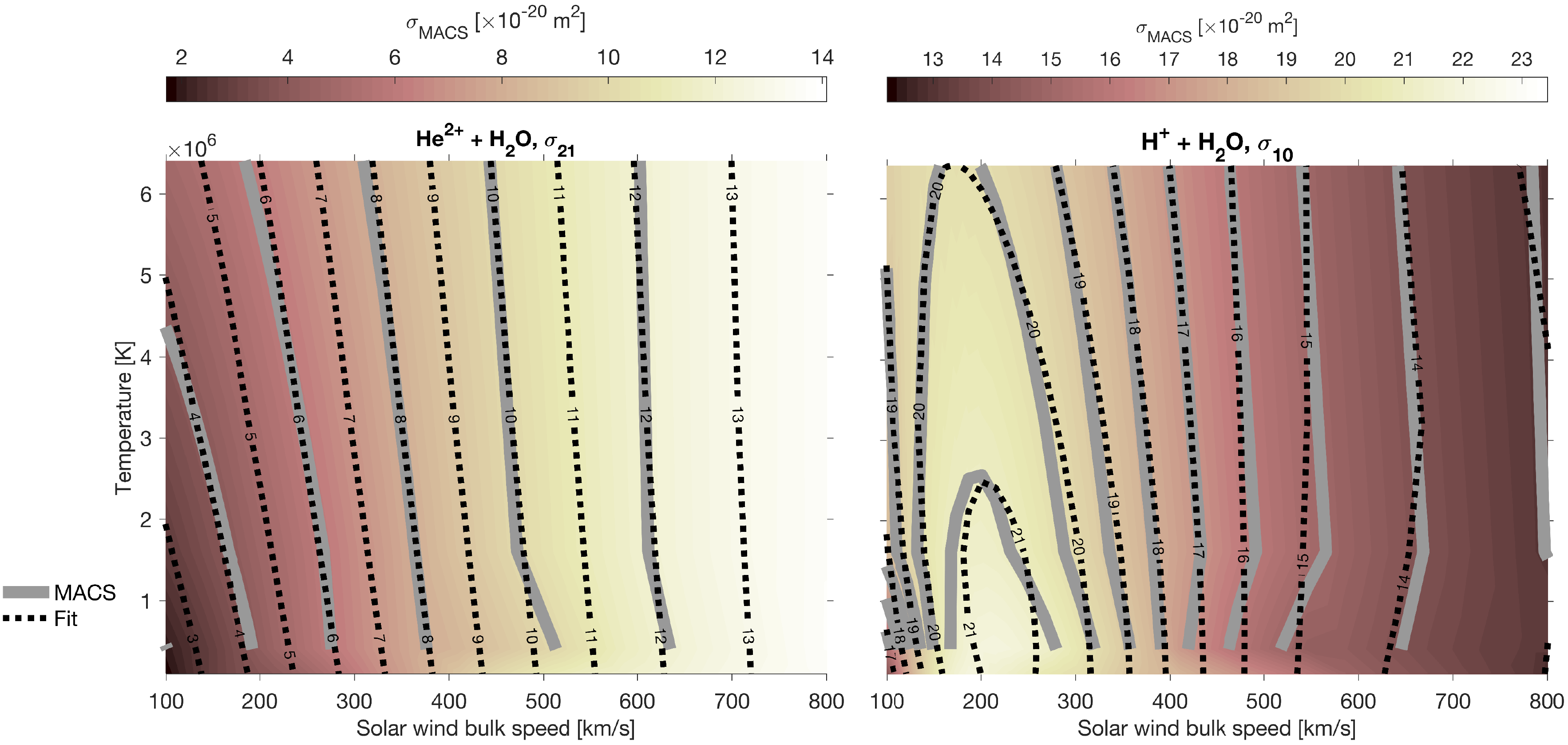}
     \caption{Maxwellian-averaged cross sections in the ($T,\varv$) plane for one-electron capture He$^{2+}-$H$_2$O (left) and H$^{+}-$H$_2$O (right). Calculated MACS contours are plotted as gray solid lines, and the corresponding fits are shown as black dotted lines.
    }
     \label{fig:crossSection_MACS_fitvalidation}
 \end{figure*}

Figure~\ref{fig:crossSection_MACS_fitvalidation} displays the comparison in the 2D $(T,\varv)$ plane between the numerically calculated single-electron capture MACS for He$^{2+}$ and H$^+$ (as gray contours) and their corresponding polynomial fits (black dotted lines). Good agreement between contours is found for helium and hydrogen, despite the discretization issue around $\varv=500$\,\kms{} for $T<10^6$\,K. Residuals become important above $800$\,\kms{}. 

For helium particles, the error with respect to the calculated charge-changing MACS was kept below $10\%$ on average for fits of degree $(4,4)$, and below $15\%$ for fits of degree $(4,5)$, all within the experimental non-averaged cross-section uncertainties. Because of rapid variations of the cross sections at low solar wind speeds, the fitting error for ionization cross sections is only $<20\%$. 
For hydrogen particles, the fitting error is correspondingly below $10\%$ for $\sigma_{10}$, $\sigma_{1-1}$ and $\sigma_{01}$, but is about $20\%$ for the other cross sections, including ionization.


\section{Summary and conclusions}

\begin{figure*}
  \includegraphics[width=\linewidth]{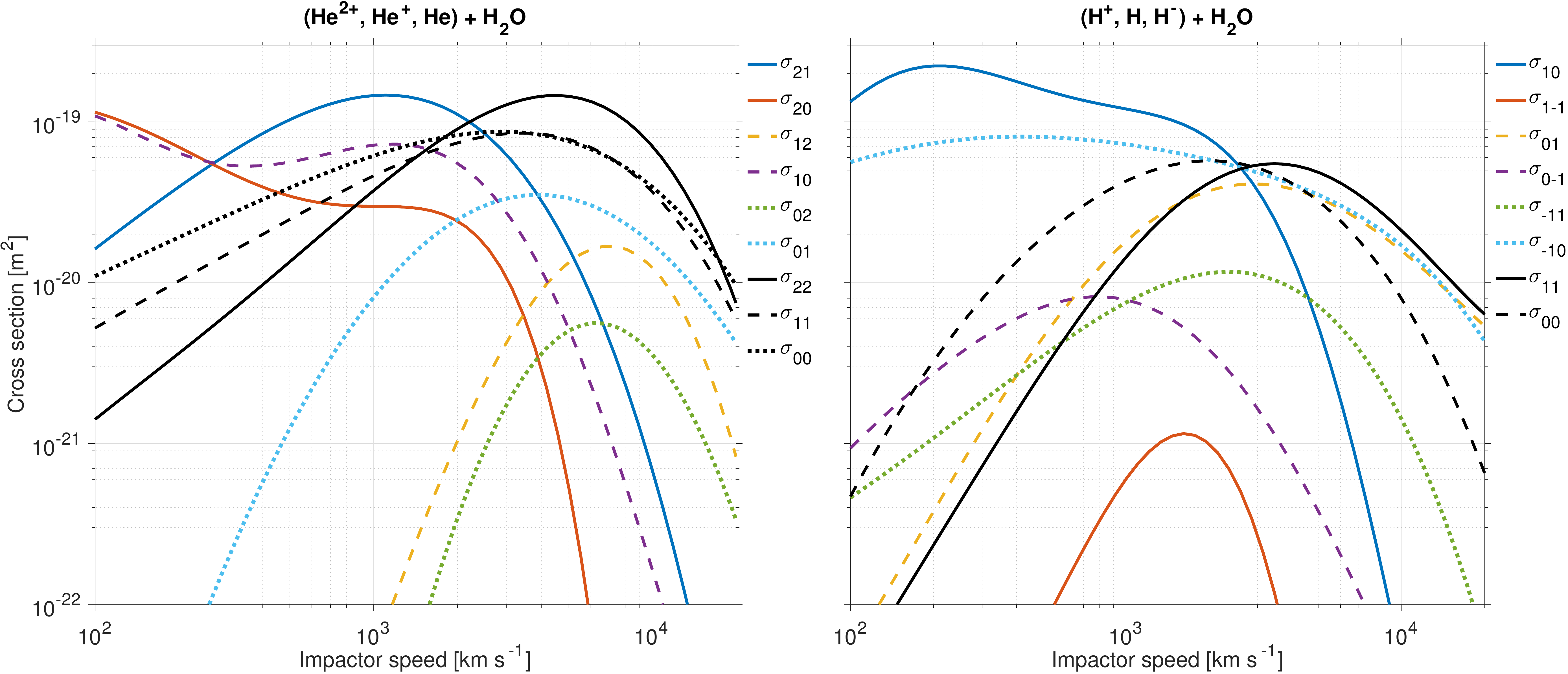}
     \caption{Recommended monochromatic charge-changing and ionization cross sections for (He$^{2+}$, He$^+$, He) and (H$^{+}$, H, H$^-$) in H$_2$O as a function of impact speed. 
     }
     \label{fig:crossSection_all}
 \end{figure*}

Figure~\ref{fig:crossSection_all} presents an overview of all fitted charge-changing and ionization cross sections for (He$^{2+}$, He$^+$, He) and (H$^{+}$, H, H$^-$) particles in a water gas, determined from a critical survey of experimental cross sections. The figure illustrates the complexity of ion-neutral interactions because the relative contribution of different reaction channels varies greatly with respect to the collision energy of the projectile. In general, electron capture (charge-exchange) reactions are the prime reactions at low velocities ($<1000$\,km/s), whereas stripping, ionization, and/or fragmentation are dominant at higher velocities.  We list our results below. 
\begin{itemize}
    \item \textbf{Helium system}. At typical solar wind velocities, the double charge-transfer reaction $\sigma_{20}$ and single-electron capture $\sigma_{10}$ dominate at low impact speeds (below $220$\,\kms{}). It is also interesting to note that their main peak is likely situated below $100$\,\kms{}, which is the limit of currently available measurements, and which would point to a semi-resonant process taking place. The contribution of double-electron loss $\sigma_{02}$ as a source of He$^{2+}$ is negligible below $1000$\,\kms{}. 
    Ionization cross sections peak at solar wind speeds above $4000$\,\kms{} where they are larger than any charge-changing cross section; they are thus expected to play only a minor role in the production of new cometary ions. Electron stripping reactions, especially $\sigma_{01}$ and to a lesser extent $\sigma_{12}$, begin to play an important role at high impact speeds (above $2000$\,\kms{}).
    
    \item \textbf{Hydrogen system}. Electron capture cross sections ($\sigma_{10}$, $\sigma_{0-1}$) usually dominate for H$^+$ and H$^0$ species for solar wind speeds below $500$\,\kms{}. For H$^-$, even though the reconstruction of electron stripping cross sections is fraught with uncertainties because of the lack of data, $\sigma_{-10}$ is expected to become the second-most important cross section after $\sigma_{10}$ at typical solar wind speeds $100-1000$\,\kms{}.
    In contrast to the case of He$^{2+}$, double charge-changing reactions (red and green lines) are negligible at typical solar wind speeds; $\sigma_{-11}$ becomes relatively important above $5000$\,\kms{}, although it remains about $\text{six}$ times lower at its peak than the other sink of H$^-$ ($\sigma_{-10}$).
    Ionization cross sections, as in the case of the helium system, tend to peak at speeds above $2000$\,\kms{}, where they constitute the main source of new ions, only rivalled there by $\sigma_{-10}$ and $\sigma_{01}$.
\end{itemize}

The velocity dependence of ion-molecule reactions implies that both the bulk velocity and the temperature of the ions need to be considered. At temperatures above $10$\,MK, the high-energy tail of the velocity distribution corresponds to the peak of the single-electron capture by He$^{2+}$ and increases the rate of single-electron capture at low solar wind speeds by almost one order of magnitude. This is also seen for ionization cross sections: MACS are amplified with respect to their parent cross sections for He and even more so for H because of the additive effect of ionization cross sections peaking at high velocities and their rapid decrease at low solar wind speeds. When cross sections peak at low impact speeds, the effect of the distribution becomes small or negligible, as is the case for the single-electron capture by H. This remark also applies to the double- and single-electron captures by He$^{2+}$ and to a lesser extent to the single-electron loss by H$^-$. This implies that shock structures found around comets and planets, where heating can reach several $10^6$\,K, may change the balance of ion production and favor processes with cross sections that peak at high energies (ionization and several charge-changing reactions), which were previously thought to be of moderate or negligible importance. 

To make the cross-section data more accurate, further experimental studies of helium and hydrogen particles in a H$_2$O gas are needed for different energy regimes and the following reactions:
\begin{itemize}
    \item \textbf{Low-energy regime} ($<100$\,eV). All $12$ charge-changing reactions for He and H particles in water, and $6$ corresponding ionization cross sections.
    \item \textbf{Medium-energy regime} ($100-10^4$\,eV). For He particles, $\sigma_{12}$, $\sigma_{02}$, $\sigma_{01}$, $\sigma_{22}$, and $\sigma_{00}$. For H particles, all cross sections, either charge-exchange or ionization.  
    \item \textbf{High-energy regime} ($10^4-10^5$\,eV). For He particles, $\sigma_{02}$, $\sigma_{01}$ , and $\sigma_{00}$. For H particles, $\sigma_{1-1}$, $\sigma_{-11}$, $\sigma_{-10}$, $\sigma_{00}$, and $\sigma_{-1-1}$. 
    \item \textbf{Very high-energy regime} ($>10^5$\,eV). For He particles, $\sigma_{12}$, $\sigma_{02}$, $\sigma_{01}$, and ionization cross sections, especially above $500$\,keV. For H particles, $\sigma_{-11}$, $\sigma_{-10}$, $\sigma_{00}$, and $\sigma_{-1-1}$.
\end{itemize}

The critical experimental survey of cross sections presented in this work has multiple applications in astrophysics and space plasma physics (other solar system bodies, H$_2$O-atmosphere exoplanets), as well as in biophysics as inputs to track-structure models \citep{Uehara2000,Uehara2002}. In parallel with this effort, a corresponding survey of existing theoretical approaches for the calculation of charge-changing and ionization cross sections, with their systematic comparison to experimental datasets, is also needed.

This article is the first part of a study on charge-exchange and ionization efficiency around comets. The second part, presented in \cite{CSW2018b} (Paper~II), develops an analytical formulation of solar wind (H$^+$, He$^{2+}$) charge exchange in cometary atmospheres using our recommended set of H$_2$O charge-exchange cross sections. The third part, presented in \cite{CSW2018c} (Paper~III), applies this analytical model and its inversions to the Rosetta Plasma Consortium datasets, and in doing so, is aimed at quantifying charge-changing reactions and comparing them to other ionization processes during the \emph{Rosetta} mission to comet 67P.

\appendix

\section{Recommended charge-changing cross sections in H$_2$O}\label{appendix:XsectionsCX}
To facilitate the book-keeping, tables of the recommended fitted monochromatic charge-changing cross sections chosen in this study are given below for the helium (table~\ref{tab:heliumXsectionsAppendix}) and hydrogen (table \ref{tab:hydrogenXsectionsAppendix}) systems. They are listed between $100$\,\kms{} and $5000$\,\kms{} ($0.05-130$\,keV/u) for all processes in numeric form, pending new experimental results. Extrapolations are indicated with an asterisk.

\begin{table*}
        \centering
        \caption{Recommended charge-changing cross sections for (He$^{2+}$, He$^+$, and He$^0$) projectiles colliding with H$_2$O vapor between $100$\,\kms{} and $5000$\,\kms{} of impact speed. 
        }
        \label{tab:heliumXsectionsAppendix}
        \begin{tabular}{llllllll} 
                \hline\hline
        Speed & Energy & \multicolumn{6}{c}{Helium-H$_2$O cross sections} \\\relax
        [\kms{}] & [keV/amu] & \multicolumn{6}{c}{[m$^2$]}  \\
         &  & $\sigma_{21}$\tablefootmark{a} & $\sigma_{20}$\tablefootmark{a} & $\sigma_{12}$\tablefootmark{a} & $\sigma_{10}$\tablefootmark{a} & $\sigma_{02}$\tablefootmark{c} & $\sigma_{01}$\tablefootmark{c} \\
         \hline
  100 & 0.052 & 1.62E-20 & 1.03E-19 & 2.80E-28$^*$ & 1.08E-19$^*$ & 9.3E-37$^*$ & 8.4E-25$^*$\\  
  150 & 0.117 & 2.74E-20 & 8.46E-20 & 2.34E-27$^*$ & 7.91E-20 & 8.0E-34$^*$ & 7.8E-24$^*$\\  
  200 & 0.209 & 3.92E-20 & 6.65E-20 & 1.06E-26$^*$ & 6.37E-20 & 6.4E-32$^*$ & 3.2E-23$^*$\\  
  250 & 0.326 & 5.12E-20 & 5.40E-20 & 3.47E-26$^*$ & 5.66E-20 & 1.5E-30$^*$ & 8.9E-23$^*$\\  
  300 & 0.470 & 6.28E-20 & 4.58E-20 & 9.12E-26$^*$ & 5.36E-20 & 1.7E-29$^*$ & 1.9E-22$^*$\\  
  350 & 0.639 & 7.39E-20 & 4.02E-20 & 2.07E-25$^*$ & 5.29E-20 & 1.2E-28$^*$ & 3.5E-22\\  
  400 & 0.835 & 8.43E-20 & 3.65E-20 & 4.19E-25$^*$ & 5.33E-20 & 5.7E-28$^*$ & 5.8E-22\\  
  450 & 1.06 & 9.39E-20 & 3.39E-20 & 7.81E-25$^*$ & 5.45E-20 & 2.2E-27$^*$ & 8.7E-22\\  
  500 & 1.30 & 1.03E-19 & 3.21E-20 & 1.36E-24$^*$ & 5.61E-20 & 6.9E-27$^*$ & 1.2E-21\\  
  550 & 1.58 & 1.10E-19 & 3.08E-20 & 2.24E-24$^*$ & 5.80E-20 & 1.9E-26$^*$ & 1.7E-21\\  
  600 & 1.88 & 1.17E-19 & 3.00E-20 & 3.52E-24$^*$ & 5.99E-20 & 4.6E-26$^*$ & 2.2E-21\\  
  700 & 2.56 & 1.29E-19 & 2.90E-20 & 7.80E-24$^*$ & 6.38E-20 & 2.0E-25$^*$ & 3.4E-21\\  
  800 & 3.34 & 1.37E-19 & 2.88E-20 & 1.54E-23$^*$ & 6.72E-20 & 6.7E-25$^*$ & 4.8E-21\\  
  900 & 4.23 & 1.42E-19 & 2.88E-20 & 2.77E-23 & 6.99E-20 & 1.8E-24$^*$ & 6.4E-21\\  
 1000 & 5.22 & 1.45E-19 & 2.90E-20 & 4.64E-23 & 7.17E-20 & 4.3E-24$^*$ & 8.1E-21\\  
 1250 & 8.16 & 1.44E-19 & 2.95E-20 & 1.34E-22 & 7.27E-20 & 2.2E-23$^*$ & 1.3E-20\\  
 1500 & 11.7 & 1.36E-19 & 2.92E-20 & 3.05E-22 & 6.94E-20 & 7.1E-23$^*$ & 1.7E-20\\  
 1750 & 16.0 & 1.24E-19 & 2.77E-20 & 5.89E-22 & 6.32E-20 & 1.7E-22$^*$ & 2.1E-20\\  
 2000 & 20.9 & 1.11E-19 & 2.52E-20 & 1.01E-21 & 5.56E-20 & 3.4E-22$^*$ & 2.5E-20\\  
 2500 & 32.6 & 8.44E-20 & 1.84E-20 & 2.29E-21 & 4.03E-20 & 9.1E-22$^*$ & 3.0E-20\\  
 3000 & 47.0 & 6.22E-20 & 1.15E-20 & 4.12E-21 & 2.76E-20 & 1.7E-21$^*$ & 3.3E-20\\  
 3500 & 63.9 & 4.50E-20 & 6.31E-21 & 6.31E-21 & 1.84E-20 & 2.7E-21$^*$ & 3.5E-20\\  
 4000 & 83.5 & 3.23E-20 & 3.11E-21 & 8.65E-21 & 1.22E-20 & 3.6E-21 & 3.5E-20\\  
 4500 & 105.7 & 2.31E-20 & 1.40E-21 & 1.09E-20 & 8.05E-21 & 4.4E-21 & 3.5E-20\\  
 5000 & 130.5 & 1.66E-20 & 5.85E-22 & 1.28E-20 & 5.34E-21 & 5.0E-21 & 3.4E-20\\
                \hline
        \end{tabular}
        \tablefoot{Energies in keV per amu are given for reference. ’E$\pm$XX’ refers to $\times 10^{\pm\textnormal{XX}}$. Values marked by an asterisk are smoothly extrapolated from the fits. Uncertainties are indicated in superscript as}
        \tablefoottext{a} $< 25\%$
        \tablefoottext{b} $25-75\%$
        \tablefoottext{c} $>75\%$.

 \end{table*}

 \begin{table*}
        \centering
        \caption{Recommended charge-changing cross sections for (H$^{+}$, H$^0$, H$^-$) projectiles colliding with H$_2$O vapor between $100$\,\kms{} and $5000$\,\kms{} of impact speed. 
        }
        \label{tab:hydrogenXsectionsAppendix}
        \begin{tabular}{llllllll} 
                \hline\hline
        Speed & Energy & \multicolumn{6}{c}{Hydrogen-H$_2$O cross sections} \\\relax
        [\kms{}] & [keV/amu] & \multicolumn{6}{c}{[m$^2$]}  \\
         &  & $\sigma_{10}$\tablefootmark{a} & $\sigma_{1-1}$\tablefootmark{c} & $\sigma_{01}$\tablefootmark{b} & $\sigma_{0-1}$\tablefootmark{c} & $\sigma_{-11}$\tablefootmark{c} & $\sigma_{-10}$\tablefootmark{c} \\
        \hline
  100 & 0.052 & 1.34E-19 & 4.8E-25 & 5.17E-23$^*$ & 9.4E-22 & 4.6E-22$^*$ & 5.7E-20\\  
  150 & 0.117 & 2.03E-19 & 1.9E-24 & 1.63E-22 & 1.8E-21 & 7.5E-22$^*$ & 6.8E-20\\  
  200 & 0.209 & 2.24E-19 & 4.3E-24 & 3.80E-22 & 2.7E-21 & 1.1E-21$^*$ & 7.4E-20\\  
  250 & 0.326 & 2.19E-19 & 8.2E-24 & 7.26E-22 & 3.6E-21 & 1.4E-21$^*$ & 7.8E-20\\  
  300 & 0.470 & 2.06E-19 & 1.4E-23 & 1.22E-21 & 4.5E-21 & 1.8E-21$^*$ & 8.0E-20\\  
  350 & 0.639 & 1.92E-19 & 2.3E-23 & 1.86E-21 & 5.3E-21 & 2.2E-21$^*$ & 8.1E-20\\  
  400 & 0.835 & 1.79E-19 & 3.5E-23 & 2.64E-21 & 6.0E-21 & 2.7E-21$^*$ & 8.2E-20\\  
  450 & 1.06 & 1.68E-19 & 5.1E-23 & 3.56E-21 & 6.6E-21 & 3.1E-21$^*$ & 8.2E-20\\  
  500 & 1.30 & 1.59E-19 & 7.2E-23 & 4.60E-21 & 7.1E-21 & 3.5E-21$^*$ & 8.1E-20\\  
  550 & 1.58 & 1.52E-19 & 9.9E-23 & 5.74E-21 & 7.4E-21 & 3.9E-21$^*$ & 8.1E-20\\  
  600 & 1.88 & 1.46E-19 & 1.3E-22 & 6.97E-21 & 7.7E-21 & 4.4E-21$^*$ & 8.0E-20\\  
  700 & 2.56 & 1.37E-19 & 2.2E-22 & 9.63E-21 & 8.1E-21 & 5.2E-21 & 7.8E-20\\  
  800 & 3.34 & 1.30E-19 & 3.3E-22 & 1.24E-20 & 8.2E-21 & 6.0E-21 & 7.7E-20\\  
  900 & 4.23 & 1.25E-19 & 4.5E-22 & 1.53E-20 & 8.0E-21 & 6.8E-21 & 7.5E-20\\  
 1000 & 5.22 & 1.21E-19 & 5.9E-22 & 1.82E-20 & 7.8E-21 & 7.5E-21 & 7.3E-20\\  
 1250 & 8.16 & 1.11E-19 & 9.2E-22 & 2.47E-20 & 6.8E-21 & 9.1E-21 & 6.9E-20\\  
 1500 & 11.7 & 1.02E-19 & 1.1E-21 & 3.02E-20 & 5.7E-21 & 1.0E-20 & 6.5E-20\\  
 1750 & 16.0 & 9.06E-20 & 1.1E-21 & 3.44E-20 & 4.7E-21 & 1.1E-20 & 6.1E-20\\  
 2000 & 20.9 & 7.91E-20 & 9.8E-22 & 3.75E-20 & 3.8E-21 & 1.1E-20 & 5.8E-20\\  
 2500 & 32.6 & 5.64E-20 & 5.7E-22 & 4.08E-20 & 2.5E-21 & 1.2E-20 & 5.3E-20\\  
 3000 & 47.0 & 3.76E-20 & 2.6E-22 & 4.15E-20 & 1.6E-21 & 1.1E-20 & 4.9E-20\\  
 3500 & 63.9 & 2.39E-20 & 1.1E-22 & 4.05E-20 & 1.1E-21 & 1.0E-20 & 4.5E-20\\  
 4000 & 83.5 & 1.47E-20 & 4.0E-23 & 3.87E-20 & 7.4E-22 & 9.3E-21 & 4.1E-20\\  
 4500 & 105.7 & 8.84E-21 & 1.5E-23 & 3.64E-20 & 5.2E-22 & 8.2E-21 & 3.8E-20\\  
 5000 & 130.5 & 5.27E-21 & 5.3E-24 & 3.39E-20 & 3.7E-22 & 7.1E-21 & 3.5E-20\\ 
                \hline
        \end{tabular}
        \tablefoot{Energies in keV per amu are given for reference. ’E$\pm$XX’ refers to $\times 10^{\pm\textnormal{XX}}$. Values marked by an asterisk are smoothly extrapolated from the fits. Uncertainties are indicated in superscript as}
        \tablefoottext{a} $< 25\%$
        \tablefoottext{b} $25-75\%$
        \tablefoottext{c} $>75\%$.
 \end{table*}

\section{Recommended ionization cross sections in H$_2$O}\label{appendix:XsectionsIon}
We present in Table~\ref{tab:XsectionsIonAppendix} our recommended fitted total monochromatic ionization cross sections for helium (He$^{2+}$, He$^+$, and He$^0$) and hydrogen (H$^{+}$ and H$^0$) particles colliding with H$_2$O. This corresponds to reactions where the impacting species does not change its charge, and is thus complementary for the net production of positive ions to the charge-changing cross sections presented in Appendix~\ref{appendix:XsectionsCX}. Ionization cross sections are listed between $100$\,\kms{} and $5000$\,\kms{} ($0.05-130$\,keV/u). Extrapolations are indicated with an asterisk.

\begin{table*}
        \centering
        \caption{Recommended total ionization cross sections for the (He$^{2+}$, He$^+$, He$^0$) and (H$^{+}$, H) systems in a H$_2$O gas between $100$\,\kms{} and $5000$\,\kms{} of impact speed. 
        }
        \label{tab:XsectionsIonAppendix}
        \begin{tabular}{lllll|ll} 
                \hline\hline
        Speed & Energy & \multicolumn{3}{c}{Helium-H$_2$O} & \multicolumn{2}{c}{Hydrogen-H$_2$O} \\\relax
        [\kms{}] & [keV/amu] & \multicolumn{3}{c}{[m$^2$]} & \multicolumn{2}{c}{[$\times 10^{-22}$\,m$^2$]}  \\
         &  & $\sigma_{22}$\tablefootmark{a} & $\sigma_{11}$\tablefootmark{a} & $\sigma_{00}$\tablefootmark{c} & $\sigma_{11}$\tablefootmark{a} & $\sigma_{00}$\tablefootmark{c} \\
         \hline
  100 & 0.052 & 1.41E-21$^*$ & 5.22E-21$^*$ & 1.1E-20$^*$ & 3.35E-23$^*$ & 4.6E-22$^*$ \\  
  150 & 0.117 & 2.46E-21$^*$ & 7.77E-21$^*$ & 1.5E-20$^*$ & 1.05E-22$^*$ & 1.5E-21 \\  
  200 & 0.209 & 3.63E-21$^*$ & 1.03E-20$^*$ & 1.9E-20$^*$ & 2.34E-22$^*$ & 3.2E-21 \\  
  250 & 0.326 & 4.95E-21$^*$ & 1.27E-20$^*$ & 2.3E-20$^*$ & 4.35E-22$^*$ & 5.4E-21 \\  
  300 & 0.470 & 6.40E-21$^*$ & 1.51E-20$^*$ & 2.6E-20$^*$ & 7.21E-22$^*$ & 7.8E-21 \\  
  350 & 0.639 & 7.99E-21$^*$ & 1.76E-20$^*$ & 3.0E-20$^*$ & 1.10E-21$^*$ & 1.1E-20 \\  
  400 & 0.835 & 9.69E-21$^*$ & 2.00E-20$^*$ & 3.3E-20$^*$ & 1.58E-21 & 1.3E-20 \\  
  450 & 1.057 & 1.15E-20$^*$ & 2.23E-20 & 3.6E-20 & 2.16E-21 & 1.6E-20 \\  
  500 & 1.305 & 1.35E-20$^*$ & 2.47E-20 & 3.9E-20 & 2.85E-21 & 1.9E-20 \\  
  550 & 1.579 & 1.55E-20$^*$ & 2.70E-20 & 4.2E-20 & 3.64E-21 & 2.2E-20 \\  
  600 & 1.879 & 1.76E-20$^*$ & 2.93E-20 & 4.4E-20 & 4.53E-21 & 2.5E-20 \\  
  700 & 2.558 & 2.22E-20$^*$ & 3.38E-20 & 4.9E-20 & 6.59E-21 & 3.0E-20 \\  
  800 & 3.341 & 2.70E-20$^*$ & 3.81E-20 & 5.4E-20 & 8.96E-21 & 3.5E-20 \\  
  900 & 4.228 & 3.20E-20$^*$ & 4.22E-20 & 5.8E-20 & 1.16E-20 & 3.9E-20 \\  
 1000 & 5.220 & 3.72E-20$^*$ & 4.61E-20 & 6.2E-20 & 1.44E-20 & 4.2E-20 \\  
 1250 & 8.156 & 5.08E-20$^*$ & 5.50E-20 & 6.9E-20 & 2.19E-20 & 4.9E-20 \\  
 1500 & 11.740 & 6.44E-20 & 6.25E-20 & 7.5E-20 & 2.93E-20 & 5.4E-20 \\  
 1750 & 15.990 & 7.76E-20 & 6.88E-20 & 8.0E-20 & 3.60E-20 & 5.6E-20 \\  
 2000 & 20.880 & 9.00E-20 & 7.38E-20 & 8.3E-20 & 4.18E-20 & 5.7E-20 \\  
 2500 & 32.620 & 1.11E-19 & 8.07E-20 & 8.6E-20 & 4.99E-20 & 5.5E-20 \\  
 3000 & 46.980 & 1.27E-19 & 8.42E-20 & 8.7E-20 & 5.40E-20 & 5.1E-20 \\  
 3500 & 63.940 & 1.38E-19 & 8.49E-20 & 8.6E-20 & 5.50E-20 & 4.6E-20 \\  
 4000 & 83.520 & 1.44E-19 & 8.38E-20 & 8.3E-20 & 5.39E-20 & 4.1E-20 \\  
 4500 & 105.700 & 1.46E-19 & 8.12E-20 & 8.0E-20 & 5.16E-20 & 3.6E-20 \\  
 5000 & 130.500 & 1.44E-19 & 7.78E-20 & 7.6E-20 & 4.85E-20 & 3.2E-20 \\  
                \hline
        \end{tabular}
        \tablefoot{Energies in keV per amu are given for reference. ’E$\pm$XX’ refers to $\times 10^{\pm\textnormal{XX}}$. Values marked by an asterisk are smoothly extrapolated from the fits. Uncertainties are indicated in superscript as}
        \tablefoottext{a} $< 25\%$
        \tablefoottext{b} $25-75\%$
        \tablefoottext{c} $>75\%$.
 \end{table*}

 \section{Recommended Maxwellian-averaged cross-section fits}\label{appendix:MACSfits}
 
 Tables~\ref{tab:heliumXsectionsMACS} and \ref{tab:hydrogenXsectionsMACS} present the recommended bivariate polynomial fit coefficients for all $12$ charge-changing cross sections for helium and hydrogen particles in water, as well as the $5$ total ionization cross sections for these two species.

\begin{table*}
        \centering
        \caption{Bivariate ($T,\varv$) polynomial fits for Maxwellian-averaged charge-changing and ionization cross sections for (He$^{2+}$, He$^+$, and He$^0$) projectiles colliding with H$_2$O vapor. The fits are valid for solar wind speeds $100-800$\,\kms{} and for solar wind temperatures ranging from $100\,000$\,K to $6.4\times10^6$\,K, unless mentioned otherwise.
        }
        \label{tab:heliumXsectionsMACS}
    \tiny
        \begin{tabular}{lrrrrrr|rrr} 
                \hline\hline
        Coefficients & \multicolumn{6}{c}{Charge-changing cross sections He-H$_2$O} & \multicolumn{3}{c}{Ionization cross sections He-H$_2$O}\\
         & $\sigma_{21}$\tablefootmark{a} & $\sigma_{20}$\tablefootmark{b} & $\sigma_{12}$\tablefootmark{a} & $\sigma_{10}$\tablefootmark{b} & $\sigma_{02}$\tablefootmark{c} & $\sigma_{01}$\tablefootmark{c} & $\sigma_{22}$\tablefootmark{a} & $\sigma_{11}$\tablefootmark{a} & $\sigma_{00}$\tablefootmark{c}\\
        \hline
        Degree $(n,m)$ & (4,4) & (4,5) & (4,4) & (4,5) & (4,4) & (4,4) & (4,4) & (4,4) & (4,4)\\
        $T$ [$\times\ 10^6$\,K] & $0.1-6.4$ & $0.1-1.6$ & $0.1-6.4$ & $0.1-1.6$ & $0.1-6.4$ & $0.1-6.4$ & $0.1-6.4$ & $0.1-6.4$  & $0.1-6.4$ \\
        \hline
$p_{00}$ &  5.634E+01 &  1.383E+03 & -9.502E-04 &  1.394E+03 &  8.597E-04 &  4.175E-01 &  4.810E+00 &  2.141E+01 &  5.216E+01 \\
$p_{10}$ &  5.115E-05 & -5.115E-04 & -5.163E-10 & -4.580E-04 & -1.461E-10 & -5.627E-07 &  7.017E-06 &  1.967E-05 &  3.431E-05 \\
$p_{01}$ &  1.440E-03 & -4.620E-03 &  1.831E-08 & -5.801E-03 & -1.421E-08 &  2.190E-07 &  5.522E-05 &  2.820E-04 &  5.829E-04 \\
$p_{20}$ & -1.716E-12 &  1.324E-10 &  1.334E-16 &  1.167E-10 & -4.114E-18 &  1.798E-13 & -9.003E-13 & -3.032E-12 & -5.332E-12 \\
$p_{11}$ & -8.517E-11 &  2.312E-09 &  4.072E-15 &  2.570E-09 &  2.078E-15 &  4.777E-12 &  7.435E-12 & -2.469E-11 & -6.996E-11 \\
$p_{02}$ &  2.829E-09 &  5.499E-09 & -8.826E-14 &  1.341E-08 &  7.786E-14 & -5.953E-11 &  5.125E-10 &  6.901E-10 &  6.584E-10 \\
$p_{30}$ &  3.962E-20 & -1.750E-17 & -2.320E-23 & -1.461E-17 &  2.075E-24 & -2.328E-20 &  1.592E-19 &  4.838E-19 &  7.900E-19 \\
$p_{21}$ &  1.890E-18 & -4.747E-16 &  8.743E-23 & -4.766E-16 & -2.851E-23 & -5.152E-19 & -1.088E-18 &  2.333E-19 &  3.801E-18 \\
$p_{12}$ &  1.859E-17 & -3.409E-15 & -1.236E-20 & -4.711E-15 & -8.113E-21 & -2.344E-18 & -2.587E-17 &  5.518E-18 &  5.026E-17 \\
$p_{03}$ & -4.679E-15 &  1.131E-15 &  1.945E-20 & -1.248E-14 & -1.747E-19 &  3.069E-16 & -1.541E-16 & -8.664E-16 & -1.216E-15 \\
$p_{40}$ & -3.331E-28 &  8.607E-25 &  2.098E-30 &  6.914E-25 & -8.472E-32 &  1.055E-27 & -9.403E-27 & -2.733E-26 & -4.221E-26 \\
$p_{31}$ & -2.015E-26 &  4.745E-23 & -5.493E-29 &  4.394E-23 & -6.737E-30 &  4.139E-26 &  4.004E-26 & -5.024E-26 & -2.637E-25 \\
$p_{22}$ & -4.579E-25 &  4.740E-22 &  8.608E-28 &  5.430E-22 &  1.719E-28 & -6.375E-26 &  7.008E-25 &  6.289E-25 & -7.978E-25 \\
$p_{13}$ &  1.647E-23 &  1.909E-21 &  2.855E-26 &  3.563E-21 &  9.351E-27 &  1.438E-24 &  1.882E-23 &  6.757E-24 & -6.629E-24 \\
$p_{04}$ &  1.810E-21 & -5.883E-21 &  4.557E-25 &  4.205E-21 &  1.396E-25 & -1.808E-22 & -7.139E-23 &  3.258E-22 &  5.335E-22 \\
$p_{41}$ & $-$ & -1.720E-30 & $-$ & -1.513E-30 & $-$ & $-$ & $-$ & $-$ & $-$ \\
$p_{32}$ & $-$ & -2.304E-29 & $-$ & -2.339E-29 & $-$ & $-$ & $-$ & $-$ & $-$ \\
$p_{23}$ & $-$ & -1.472E-28 & $-$ & -1.980E-28 & $-$ & $-$ & $-$ & $-$ & $-$ \\
$p_{14}$ & $-$ & -2.730E-28 & $-$ & -9.401E-28 & $-$ & $-$ & $-$ & $-$ & $-$ \\
$p_{05}$ & $-$ &  2.774E-27 & $-$ &  3.938E-30 & $-$ & $-$ & $-$ & $-$ & $-$ \\
         \hline
        \end{tabular}
        \tablefoot{'E$\pm$XX' refers to $\times10^{\pm\textnormal{XX}}$. To obtain a better fit, the maximum temperature for the fit was reduced to $1.6\times10^6$\,K for $\sigma_{20}$ and $\sigma_{10}$. All polynomial fits should be multiplied by $10^{-22}$ to scale to the final cross section. Consequently, the (4,4) polynomial model is
        \begin{align*}
            \sigma_\textnormal{MACS}(T,\varv)  = 10^{-22} \left(p_{00} + p_{10} T + p_{01} \varv + p_{20} T^2 + p_{11} T \varv + p_{02} \varv^2 + p_{30} T^3 + p_{21} T^2 \varv  + p_{12} T \varv^2+ p_{03} \varv^3 + p_{40} T^4 + p_{31} T^3 \varv + p_{22} T^2 \varv^2 + p_{13} T \varv^3 + p_{04} \varv^4\right)
        \end{align*}
        , with $\sigma_\textnormal{MACS}$ expressed in m$^2$, $T$ in K and $\varv$ in m\,s$^{-1}$. The (4,5) polynomial model takes the following form:
        \begin{align*}
            \sigma_\textnormal{MACS}(T,\varv)  = 10^{-22} &\left(p_{00} + p_{10} T + p_{01} \varv + p_{20} T^2 + p_{11} T \varv + p_{02} \varv^2 + p_{30} T^3 
                    + p_{21} T^2 \varv + p_{12} T \varv^2 + p_{03} \varv^3 + p_{40} T^4 + p_{31} T^3 \varv 
                    + p_{22} T^2 \varv^2 + p_{13} T \varv^3 + p_{04} \varv^4 \right.\\
                    &\left. + p_{41} T^4 \varv + p_{32} T^3 \varv^2 
                    + p_{23} T^2 \varv^3 + p_{14} T \varv^4 + p_{05} \varv^5\right)
        \end{align*}
        . Errors are a combination of those of the non-averaged cross sections and that of the bivariate fits. They are indicated in superscript as}
        \tablefoottext{a} $< 25\%$
        \tablefoottext{b} $25-75\%$
        \tablefoottext{c} $>75\%$.
\end{table*}

\begin{table*}
        \centering
        \caption{Bivariate ($T,\varv$) polynomial fits for Maxwellian-averaged charge-changing and ionization cross sections for (H$^{+}$, H$^0$, and H$^-$) projectiles colliding with H$_2$O vapor. The fits are valid for solar wind speeds $100-800$\,\kms{} and for solar wind temperatures ranging from $25\,000$\,K to $10\times10^6$\,K, unless mentioned otherwise. 
        }
        \label{tab:hydrogenXsectionsMACS}
    \tiny
        \begin{tabular}{lrrrrrr|rr} 
                \hline\hline
        Coefficients & \multicolumn{6}{c}{Charge-changing cross sections H-H$_2$O} & \multicolumn{2}{c}{Ionization cross sections H-H$_2$O}\\
        & $\sigma_{10}$\tablefootmark{a} & $\sigma_{1-1}$\tablefootmark{c} & $\sigma_{01}$\tablefootmark{b} & $\sigma_{0-1}$\tablefootmark{c} & $\sigma_{-11}$\tablefootmark{c} & $\sigma_{-10}$\tablefootmark{c} & $\sigma_{11}$\tablefootmark{a} & $\sigma_{00}$\tablefootmark{a}\\
        Degree $(n,m)$ & (4,5) & (4,4) & (4,4) & (4,5) & (4,4) & (4,5) & (4,4) & (4,4)\\
        $T$ [$\times\ 10^6$\,K] & $0.025-10$ & $0.025-1.6$ & $0.025-10$ & $0.025-1.6$ & $0.025-10$ & $0.025-10$ & $0.025-10$ & $0.025-10$ \\
        \hline
$p_{00}$ &  6.151E+01 & -3.085E-02 & -2.149E+00 &  1.899E+00 &  1.679E+00 &  5.445E+02 & -9.696E-01 & -1.212E+01 \\
$p_{10}$ &  3.886E-04 &  2.826E-08 &  5.952E-06 &  2.109E-05 &  6.312E-06 &  1.003E-04 &  3.118E-06 &  3.543E-05 \\
$p_{01}$ &  2.331E-02 &  5.110E-07 & -1.595E-05 &  5.904E-05 &  2.749E-05 &  1.249E-03 & -6.480E-06 &  7.489E-05 \\
$p_{20}$ & -5.076E-11 &  3.080E-14 &  7.320E-14 & -4.450E-12 & -7.040E-13 & -2.007E-11 &  2.893E-13 & -3.656E-12 \\
$p_{11}$ & -3.137E-09 &  1.077E-13 &  3.381E-12 & -4.133E-11 & -6.877E-12 & -4.151E-10 &  3.639E-12 & -2.506E-11 \\
$p_{02}$ & -9.085E-08 & -3.405E-12 &  2.546E-10 &  5.611E-10 &  1.342E-10 & -1.057E-09 &  9.705E-11 &  1.133E-09 \\
$p_{30}$ &  4.472E-18 & -3.201E-21 & -4.963E-21 &  4.808E-19 &  6.839E-20 &  1.948E-18 & -2.719E-20 &  3.602E-19 \\
$p_{21}$ &  2.028E-16 & -1.941E-20 & -1.956E-19 &  1.057E-17 &  6.270E-19 &  6.146E-17 & -3.573E-19 &  2.821E-18 \\
$p_{12}$ &  8.870E-15 &  8.670E-19 & -9.357E-19 & -3.972E-17 &  1.168E-18 &  4.984E-16 &  6.289E-18 & -3.051E-17 \\
$p_{03}$ &  1.455E-13 &  1.389E-17 & -2.668E-17 & -1.115E-15 & -1.373E-16 & -2.555E-15 &  1.024E-16 & -1.098E-15 \\
$p_{40}$ & -1.530E-25 &  1.186E-28 &  1.190E-28 & -1.797E-26 & -2.416E-27 & -6.820E-26 &  9.413E-28 & -1.268E-26 \\
$p_{31}$ & -1.172E-23 &  9.531E-28 &  7.201E-27 & -1.087E-24 & -2.645E-26 & -4.878E-24 &  1.426E-26 & -1.379E-25 \\
$p_{22}$ & -2.677E-22 &  1.022E-27 &  7.523E-26 & -1.775E-24 & -1.521E-25 & -4.111E-23 &  1.166E-25 & -2.291E-25 \\
$p_{13}$ & -1.050E-20 & -9.332E-25 & -6.307E-24 &  8.836E-23 &  4.486E-25 & -2.427E-22 & -9.814E-24 &  2.507E-23 \\
$p_{04}$ & -1.030E-19 & -4.447E-24 & -1.437E-23 &  7.177E-22 &  5.356E-23 &  4.788E-21 & -3.734E-23 &  3.310E-22 \\
$p_{41}$ &  2.810E-31 & $-$ & $-$ &  3.524E-32 & $-$ &  1.357E-31 & $-$ & $-$ \\
$p_{32}$ &  6.353E-30 & $-$ & $-$ &  2.508E-31 & $-$ &  1.766E-30 & $-$ & $-$ \\
$p_{23}$ &  1.181E-28 & $-$ & $-$ & -2.311E-30 & $-$ &  5.386E-30 & $-$ & $-$ \\
$p_{14}$ &  4.464E-27 & $-$ & $-$ & -3.188E-29 & $-$ &  4.929E-29 & $-$ & $-$ \\
$p_{05}$ &  2.572E-26 & $-$ & $-$ & -1.419E-28 & $-$ & -2.276E-27 & $-$ & $-$ \\
         \hline
        \end{tabular}
        \tablefoot{'E$\pm$XX' refers to $\times10^{\pm\textnormal{XX}}$. Bivariate polynomial fits are given in the Notes of Table~\ref{tab:heliumXsectionsMACS}.
        Errors are a combination of those of the non-averaged cross sections and that of the bivariate fits. They are indicated in superscript as}
        \tablefoottext{a} $< 25\%$
        \tablefoottext{b} $25-75\%$
        \tablefoottext{c} $>75\%$.
\end{table*}

\begin{acknowledgements}
The work at University of Oslo was funded by the Norwegian Research Council grant No.~240\,000. The work at NASA/SSAI was supported by NASA Astrobiology Institute grant NNX15AE05G and by the NASA HIDEE Program. Work at the Royal Belgian Institute for Space Aeronomy was supported by the Belgian Science Policy Office through the Solar-Terrestrial Centre of Excellence. Work at Ume\aa{} University was funded by SNSB grant 201/15 and SNSA grant 108/18. Work at Imperial College London was supported by STFC of UK under grant ST/N000692/1 and by ESA under contract No. 4000119035/16/ES/JD.
C.S.W. is grateful to S. Barabash (IRF Kiruna, Sweden) for suggesting to investigate electron stripping processes at a comet. The authors thank the ISSI International Team "Plasma Environment of comet 67P after \emph{Rosetta}" for fruitful discussions and collaborations. C.S.W. acknowledges the \emph{Cross sections in ion-atom collisions} (\url{http://cdfe.sinp.msu.ru/services/cccs/HTM/main.htm}) database hosted by the Skobeltsyn Institute of Nuclear Physics, Lomonosov Moscow State University (Russia) and maintained by N.V. Novikov and Ya. A. Teplova, which was instrumental in finding cross-section datasets for H$_2$, H, O$_2$ , and O targets. C.S.W. is indebted to M.S.W. and L.S.W. for cynosural advice throughout this obstinate quest for the best experimental cross sections. 
Datasets of the \emph{Rosetta} mission can be freely accessed from ESA's Planetary Science Archive (\url{http://archives.esac.esa.int/psa}). The authors thank Astrid Peter and the A\&A editing team for help in preparing our series of three manuscripts on cometary charge-changing processes.
\end{acknowledgements}


\bibliographystyle{aa}
\bibliography{references} 

\begin{thebibliography}{105}
\expandafter\ifx\csname natexlab\endcsname\relax\def\natexlab#1{#1}\fi

\bibitem[{{Allison}(1958)}]{Allison1958}
{Allison}, S.~K. 1958, Rev. Mod. Phys., 30, 1137

\bibitem[{{Alvarado} {et~al.}(2005){Alvarado}, {Hoekstra}, \&
  {Schlath{\"o}lter}}]{Alvarado2005}
{Alvarado}, F., {Hoekstra}, R., \& {Schlath{\"o}lter}, T. 2005, J. Phys. B:
  Atom. Mol. and Opt. Phys., 38, 4085

\bibitem[{{Bailey} \& {Mahadevan}(1970)}]{Bailey1970}
{Bailey}, T.~L. \& {Mahadevan}, P. 1970, J. Chem. Phys., 52, 179

\bibitem[{{Banks} \& {Kockarts}(1973)}]{Banks1973a}
{Banks}, P.~M. \& {Kockarts}, G. 1973, {Aeronomy, Part A} (Academic Press, New
  York and London)

\bibitem[{{Baribaud}(1972)}]{Baribaud1972}
{Baribaud}, M. 1972, PhD Thesis, Universit\'{e} de Grenoble

\bibitem[{{Baribaud} {et~al.}(1971){Baribaud}, {Monte}, \&
  {Zadworny}}]{Baribaud1971}
{Baribaud}, M., {Monte}, J., \& {Zadworny}, F. 1971, Compt. Rend. Acad. Sci.
  Paris, 272B, 457

\bibitem[{{Barnett} {et~al.}(1990){Barnett}, {Hunter}, {Fitzpatrick},
  {Alvarez}, {Cisneros}, \& {Phaneuf}}]{Barnett1990}
{Barnett}, C.~F., {Hunter}, H.~T., {Fitzpatrick}, M.~I., {et~al.} 1990, NASA
  STI/Recon Technical Report N, 91, 13238

\bibitem[{{Barnett} {et~al.}(1977){Barnett}, {Ray}, {Ricci}, {Wilker},
  {McDaniel}, {Thomas}, \& {Gilbody}}]{Barnett1977}
{Barnett}, C.~F., {Ray}, J.~A., {Ricci}, E., {et~al.} 1977, {Atomic data for
  controlled fusion research}, Tech. rep., ORNL

\bibitem[{{Behar} {et~al.}(2017){Behar}, {Nilsson}, {Alho}, {Goetz}, \&
  {Tsurutani}}]{Behar2017}
{Behar}, E., {Nilsson}, H., {Alho}, M., {Goetz}, C., \& {Tsurutani}, B. 2017,
  Month. Not. Roy. Astron. Soc., 469, S396

\bibitem[{{Belki{\'c}} {et~al.}(1979){Belki{\'c}}, {Gayet}, \&
  {Salin}}]{Belkic1979}
{Belki{\'c}}, D., {Gayet}, R., \& {Salin}, A. 1979, \physrep, 56, 279

\bibitem[{{Berger} {et~al.}(1993){Berger}, {Inokuti}, {Andersen}, {Bichsel},
  {Powers}, {Seltzer}, {Thwaites}, \& {Watt}}]{ICRU1993}
{Berger}, M.~J., {Inokuti}, M., {Andersen}, H.~H., {et~al.} 1993, J. Int. Comm.
  Rad. Units Meas., os25, NP

\bibitem[{Berkner {et~al.}(1970)Berkner, Pyle, \& Stearns}]{Berkner1970}
Berkner, K., Pyle, R., \& Stearns, J. 1970, Nuclear Fusion, 10, 145

\bibitem[{{Bissinger} {et~al.}(1982){Bissinger}, {Joyce}, {Lapicki}, {Laubert},
  \& {Varghese}}]{Bissinger1982}
{Bissinger}, G., {Joyce}, J.~M., {Lapicki}, G., {Laubert}, R., \& {Varghese},
  S.~L. 1982, Phys. Rev. Lett., 49, 318

\bibitem[{{Bodewits} {et~al.}(2007){Bodewits}, {Christian}, {Torney}, {Dryer},
  {Lisse}, {Dennerl}, {Zurbuchen}, {Wolk}, {Tielens}, \&
  {Hoekstra}}]{Bodewits2007AA}
{Bodewits}, D., {Christian}, D.~J., {Torney}, M., {et~al.} 2007, A\&A, 469,
  1183

\bibitem[{{Bodewits} {et~al.}(2006){Bodewits}, {Hoekstra}, {Seredyuk},
  {McCullough}, {Jones}, \& {Tielens}}]{Bodewits2006}
{Bodewits}, D., {Hoekstra}, R., {Seredyuk}, B., {et~al.} 2006, Astrophys. J.,
  642, 593

\bibitem[{{Bodewits} {et~al.}(2004){Bodewits}, {Juh{\'a}sz}, {Hoekstra}, \&
  {Tielens}}]{Bodewits2004ApJ}
{Bodewits}, D., {Juh{\'a}sz}, Z., {Hoekstra}, R., \& {Tielens}, A.~G.~G.~M.
  2004, Astrophys. J., 606, L81

\bibitem[{{Bolorizadeh} \& {Rudd}(1986{\natexlab{a}})}]{Bolorizadeh1986_Hp}
{Bolorizadeh}, M.~A. \& {Rudd}, M.~E. 1986{\natexlab{a}}, Phys. Rev. A, 33, 888

\bibitem[{{Bolorizadeh} \& {Rudd}(1986{\natexlab{b}})}]{Bolorizadeh1986_H}
{Bolorizadeh}, M.~A. \& {Rudd}, M.~E. 1986{\natexlab{b}}, Phys. Rev. A, 33, 893

\bibitem[{{Bragg} \& {Kleeman}(1905)}]{Bragg1905}
{Bragg}, W.~H. \& {Kleeman}, R. 1905, Lond. Edinb. Dubl. Phil. Mag. J. Sci.,
  10, 318

\bibitem[{{Bryan} {et~al.}(1990){Bryan}, {Freeman}, \& {Monce}}]{Bryan1990}
{Bryan}, E.~L., {Freeman}, E.~J., \& {Monce}, M.~N. 1990, Phys. Rev. A, 42,
  6423

\bibitem[{{Cable}(1970)}]{Cable1970}
{Cable}, P. 1970, PhD Thesis, University of Maryland

\bibitem[{{Combi} {et~al.}(2004){Combi}, {Harris}, \& {Smyth}}]{Combi2004}
{Combi}, M.~R., {Harris}, W.~M., \& {Smyth}, W.~H. 2004, in Comets II, ed.
  M.~C. {Festou}, H.~U. {Keller}, \& H.~A. {Weaver} (1510 E. University Blvd.,
  P.O. Box 210055, Tucson, AZ 85721-0055: The University of Arizona Press),
  523--552

\bibitem[{{Coplan} \& {Ogilvie}(1970)}]{Coplan1970}
{Coplan}, M.~A. \& {Ogilvie}, K.~W. 1970, J. Chem. Phys., 52, 4154

\bibitem[{{Cravens}(1997)}]{Cravens1997}
{Cravens}, T.~E. 1997, Geophys. Res. Lett., 24, 105

\bibitem[{{Dagnac} {et~al.}(1969){Dagnac}, {Blanc}, {Kucuk}, \&
  {Molina}}]{Dagnac1969}
{Dagnac}, R., {Blanc}, D., {Kucuk}, M., \& {Molina}, D. 1969, Compt. Rend.
  Acad. Sci. Paris, 268B, 676

\bibitem[{{Dagnac} {et~al.}(1970){Dagnac}, {Blanc}, \& {Molina}}]{Dagnac1970}
{Dagnac}, R., {Blanc}, D., \& {Molina}, D. 1970, J. Phys. B: At. Mol. Phys., 3,
  1239

\bibitem[{{Dalgarno}(1958)}]{Dalgarno1958}
{Dalgarno}, A. 1958, Phil. Trans. Roy. Soc. London Series A, 250, 426

\bibitem[{{Dennerl}(2010)}]{Dennerl2010}
{Dennerl}, K. 2010, Space Sci. Rev., 157, 57

\bibitem[{{Dingfelder} {et~al.}(2000){Dingfelder}, {Inokuti}, \&
  {Paretzke}}]{Dingfelder2000}
{Dingfelder}, M., {Inokuti}, M., \& {Paretzke}, H.~G. 2000, Rad. Phys. Chem.,
  59, 255

\bibitem[{{Endo} {et~al.}(2002){Endo}, {Yoshida}, {Nikjoo}, {Uehara}, {Hoshi},
  {Ishikawa}, \& {Shizuma}}]{Endo2002}
{Endo}, S., {Yoshida}, E., {Nikjoo}, H., {et~al.} 2002, Nucl. Inst. Meth. Phys.
  Res. B, 194, 123

\bibitem[{{Fogel} {et~al.}(1957){Fogel}, {Ankudinov}, \&
  {Slabospitskii}}]{Fogel1957}
{Fogel}, I.~M., {Ankudinov}, V.~A., \& {Slabospitskii}, R.~E. 1957, Soviet J.
  Exp. Theo. Phys., 5, 382

\bibitem[{{Fuselier} {et~al.}(1991){Fuselier}, {Shelley}, {Goldstein},
  {Goldstein}, {Neugebauer}, {Ip}, {Balsiger}, \& {Reme}}]{Fuselier1991}
{Fuselier}, S.~A., {Shelley}, E.~G., {Goldstein}, B.~E., {et~al.} 1991, \apj,
  379, 734

\bibitem[{{Gealy} \& {van Zyl}(1987{\natexlab{a}})}]{Gealy1987_Hp}
{Gealy}, M.~W. \& {van Zyl}, B. 1987{\natexlab{a}}, Phys. Rev. A, 36, 3091

\bibitem[{{Gealy} \& {van Zyl}(1987{\natexlab{b}})}]{Gealy1987_H}
{Gealy}, M.~W. \& {van Zyl}, B. 1987{\natexlab{b}}, Phys. Rev. A, 36, 3100

\bibitem[{{Geddes} {et~al.}(1980){Geddes}, {Hill}, {Shah}, {Goffe}, \&
  {Gilbody}}]{Geddes1980}
{Geddes}, J., {Hill}, J., {Shah}, M.~B., {Goffe}, T.~V., \& {Gilbody}, H.~B.
  1980, J. Phys. B At. Mol. Phys., 13, 319

\bibitem[{{Gobet} {et~al.}(2004){Gobet}, {Eden}, {Coupier}, {Tabet}, {Farizon},
  {Farizon}, {Gaillard}, {Carr{\'e}}, {Ouaskit}, {M{\"a}rk}, \&
  {Scheier}}]{Gobet2004}
{Gobet}, F., {Eden}, S., {Coupier}, B., {et~al.} 2004, Phys. Rev. A, 70, 062716

\bibitem[{{Gobet} {et~al.}(2006){Gobet}, {Eden}, {Coupier}, {Tabet}, {Farizon},
  {Farizon}, {Gaillard}, {Ouaskit}, {Carr{\'e}}, \& {M{\"a}rk}}]{Gobet2006}
{Gobet}, F., {Eden}, S., {Coupier}, B., {et~al.} 2006, Chem. Phys. Lett., 421,
  68

\bibitem[{{Gobet} {et~al.}(2001){Gobet}, {Farizon}, {Farizon}, {Gaillard},
  {Carr{\'e}}, {Lezius}, {Scheier}, \& {M{\"a}rk}}]{Gobet2001}
{Gobet}, F., {Farizon}, B., {Farizon}, M., {et~al.} 2001, Phys. Rev. Lett., 86,
  3751

\bibitem[{{Green} \& {McNeal}(1971)}]{Green1971}
{Green}, A.~E.~S. \& {McNeal}, R.~J. 1971, \jgr, 76, 133

\bibitem[{{Greenwood} {et~al.}(2000){Greenwood}, {Chutjian}, \&
  {Smith}}]{Greenwood2000}
{Greenwood}, J.~B., {Chutjian}, A., \& {Smith}, S.~J. 2000, Astrophys. J., 529,
  605

\bibitem[{{Greenwood} {et~al.}(2004){Greenwood}, {Mawhorter}, {Cadez},
  {Lozano}, {Smith}, \& {Chutjian}}]{Greenwood2004}
{Greenwood}, J.~B., {Mawhorter}, R.~J., {Cadez}, I., {et~al.} 2004, Phys.
  Script. T, 110, 358

\bibitem[{{Gunell} {et~al.}(2018){Gunell}, {Goetz}, {Simon Wedlund},
  {Lindkvist}, {Hamrin}, {Nilsson}, {Llera}, {Eriksson}, \&
  {Holmstr{\"o}m}}]{Gunell2018}
{Gunell}, H., {Goetz}, C., {Simon Wedlund}, C., {et~al.} 2018, \aap, 619, L2

\bibitem[{{Hill} {et~al.}(1979){Hill}, {Geddes}, \& {Gilbody}}]{Hill1979}
{Hill}, J., {Geddes}, J., \& {Gilbody}, H.~B. 1979, J. Phys. B At. Mol. Phys.,
  12, 3341

\bibitem[{{Hoekstra} {et~al.}(1998){Hoekstra}, {Anderson}, {Bliek}, {von
  Hellermann}, {Maggi}, {Olson}, \& {Summers}}]{Hoekstra1998}
{Hoekstra}, R., {Anderson}, H., {Bliek}, F.~W., {et~al.} 1998, Plasma Phys.
  Control. Fus., 40, 1541

\bibitem[{{Hoekstra} {et~al.}(2006){Hoekstra}, {Bodewits}, {Knoop},
  {Morgenstern}, {Mendez}, {Errea}, {Illecas}, {Macias}, {Pons}, {Riera},
  {Aymayr}, \& {Winter}}]{Hoekstra2006}
{Hoekstra}, R., {Bodewits}, D., {Knoop}, S., {et~al.} 2006, {Charge exchange
  data for alpha particles interacting with atoms and molecules}, IAEA Atomic
  Plasma-Material Interaction Data for Fusion No.~13 (Vienna: International
  Atomic Energy Agency), 8--20

\bibitem[{{Huq} {et~al.}(1983){Huq}, {Doverspike}, \&
  {Champion}}]{Huq1983elecdetach}
{Huq}, M.~S., {Doverspike}, L.~D., \& {Champion}, R.~L. 1983, Phys. Rev. A, 27,
  2831

\bibitem[{{Hvelplund} \& {Andersen}(1982)}]{Hvelplund1982elecloss}
{Hvelplund}, P. \& {Andersen}, A. 1982, Phys. Script., 26, 370

\bibitem[{{Ip}(1989)}]{Ip1989}
{Ip}, W.-H. 1989, Astrophys. J., 343, 946

\bibitem[{{Isler}(1977)}]{Isler1977}
{Isler}, R.~C. 1977, Phys. Rev. Lett., 38, 1359

\bibitem[{{Itikawa} \& {Mason}(2005)}]{Itikawa2005}
{Itikawa}, Y. \& {Mason}, N. 2005, J. Phys. Chem. Ref. Data, 34, 1

\bibitem[{{Itoh} {et~al.}(1980{\natexlab{a}}){Itoh}, {Asari}, \&
  {Fukuzawa}}]{Itoh1980capture}
{Itoh}, A., {Asari}, M., \& {Fukuzawa}, F. 1980{\natexlab{a}}, J. Phys. Soc.
  Japan, 48, 943

\bibitem[{{Itoh} {et~al.}(1980{\natexlab{b}}){Itoh}, {Ohnishi}, \&
  {Fukuzawa}}]{Itoh1980loss}
{Itoh}, A., {Ohnishi}, K., \& {Fukuzawa}, F. 1980{\natexlab{b}}, J. Phys. Soc.
  Japan, 49, 1513

\bibitem[{{Kharchenko} {et~al.}(2003){Kharchenko}, {Rigazio}, {Dalgarno}, \&
  {Krasnopolsky}}]{Kharchenko2003}
{Kharchenko}, V., {Rigazio}, M., {Dalgarno}, A., \& {Krasnopolsky}, V.~A. 2003,
  Astrophys. J., 585, L73

\bibitem[{{Koenders} {et~al.}(2013){Koenders}, {Glassmeier}, {Richter},
  {Motschmann}, \& {Rubin}}]{Koenders2013}
{Koenders}, C., {Glassmeier}, K.-H., {Richter}, I., {Motschmann}, U., \&
  {Rubin}, M. 2013, Plan. Space Sci., 87, 85

\bibitem[{{Koopman}(1968)}]{Koopman1968}
{Koopman}, D.~W. 1968, Phys. Rev., 166, 57

\bibitem[{{Kozlov} \& {Bondar'}(1966)}]{Kozlov1966}
{Kozlov}, V.~F. \& {Bondar'}, S.~A. 1966, Soviet J. Exp. Theo. Phys., 23, 195

\bibitem[{{L\"{a}uter} {et~al.}(2019){L\"{a}uter}, {Kramer}, {Rubin}, \&
  {Altwegg}}]{Lauter2018}
{L\"{a}uter}, M., {Kramer}, T., {Rubin}, M., \& {Altwegg}, K. 2019, Month. Not.
  Roy. Astron. Soc., 483, 852

\bibitem[{{Lichtenberg} {et~al.}(1980){Lichtenberg}, {Bethge}, \&
  {Schmidt-Bocking}}]{Lichtenberg1980}
{Lichtenberg}, W.~J., {Bethge}, K., \& {Schmidt-Bocking}, H. 1980, J. Phys. B
  At. Mol. Phys., 13, 343

\bibitem[{{Lindsay} {et~al.}(1997){Lindsay}, {Sieglaff}, {Smith}, \&
  {Stebbings}}]{Lindsay1997}
{Lindsay}, B.~G., {Sieglaff}, D.~R., {Smith}, K.~A., \& {Stebbings}, R.~F.
  1997, Phys. Rev. A, 55, 3945

\bibitem[{{Lindsay} \& {Stebbings}(2005)}]{Lindsay2005}
{Lindsay}, B.~G. \& {Stebbings}, R.~F. 2005, J. Geophys. Res., 110, A12213

\bibitem[{{Lisse} {et~al.}(1996){Lisse}, {Dennerl}, {Englhauser}, {Harden},
  {Marshall}, {Mumma}, {Petre}, {Pye}, {Ricketts}, {Schmitt}, {Trumper}, \&
  {West}}]{Lisse1996}
{Lisse}, C.~M., {Dennerl}, K., {Englhauser}, J., {et~al.} 1996, Science, 274,
  205

\bibitem[{{Livadiotis} {et~al.}(2018){Livadiotis}, {Desai}, \& {Wilson
  III}}]{Livadiotis2018}
{Livadiotis}, G., {Desai}, M.~I., \& {Wilson III}, L.~B. 2018, \apj, 853, 142

\bibitem[{{Luna} {et~al.}(2007){Luna}, {de Barros}, {Wyer}, {Scully},
  {Lecointre}, {Garcia}, {Sigaud}, {Santos}, {Senthil}, {Shah}, {Latimer}, \&
  {Montenegro}}]{Luna2007}
{Luna}, H., {de Barros}, A.~L.~F., {Wyer}, J.~A., {et~al.} 2007, Phys. Rev. A,
  75, 042711

\bibitem[{{Mada} {et~al.}(2007){Mada}, {Hida}, {Kimura}, {Pichl}, {Liebermann},
  {Li}, \& {Buenker}}]{Mada2007}
{Mada}, S., {Hida}, K.-N., {Kimura}, M., {et~al.} 2007, Phys. Rev. A, 75,
  022706

\bibitem[{{Mahadevan} \& {Magnuson}(1968)}]{Mahadevan1968}
{Mahadevan}, P. \& {Magnuson}, G.~D. 1968, Phys. Rev., 171, 103

\bibitem[{{McClure}(1963)}]{McClure1963}
{McClure}, G.~W. 1963, Phys. Rev., 132, 1636

\bibitem[{{McNeal} \& {Birely}(1973)}]{McNeal1973}
{McNeal}, R.~J. \& {Birely}, J.~H. 1973, Rev. Geophys. Space Phys., 11, 633

\bibitem[{{Meyer-Vernet}(2012)}]{Meyer2012}
{Meyer-Vernet}, N. 2012, {Basics of the Solar Wind} (Cambridge, UK: Cambridge
  University Press)

\bibitem[{{Mullen} {et~al.}(2017){Mullen}, {Cumbee}, {Lyons}, {Gu}, {Kaastra},
  {Shelton}, \& {Stancil}}]{Mullen2017}
{Mullen}, P.~D., {Cumbee}, R.~S., {Lyons}, D., {et~al.} 2017, \apj, 844, 7

\bibitem[{{Nikjoo} {et~al.}(2012){Nikjoo}, {Uehara}, \&
  {Emfietzoglou}}]{Nikjoo2012}
{Nikjoo}, H., {Uehara}, S., \& {Emfietzoglou}, D. 2012, Interaction of
  Radiation with Matter (CRC Press)

\bibitem[{{Nilsson} {et~al.}(2015){Nilsson}, {Stenberg Wieser}, {Behar}, {Simon
  Wedlund}, {Gunell}, {Yamauchi}, {Lundin}, {Barabash}, {Wieser}, {Carr},
  {Cupido}, {Burch}, {Fedorov}, {Sauvaud}, {Koskinen}, {Kallio}, {Lebreton},
  {Eriksson}, {Edberg}, {Goldstein}, {Henri}, {Koenders}, {Mokashi}, {Nemeth},
  {Richter}, {Szego}, {Volwerk}, {Vallat}, \& {Rubin}}]{Nilsson2015}
{Nilsson}, H., {Stenberg Wieser}, G., {Behar}, E., {et~al.} 2015, Science, 347,
  571

\bibitem[{{Novikov} \& {Teplova}(2009)}]{Novikov2009}
{Novikov}, N.~V. \& {Teplova}, Y.~A. 2009, in J. Phys. Conf. Ser., Vol. 194, J.
  Phys. Conf. Ser., 082032

\bibitem[{{Phelps}(1990)}]{Phelps1990}
{Phelps}, A.~V. 1990, J. Phys. Chem. Ref. Data, 19, 653

\bibitem[{{Risley} \& {Geballe}(1974)}]{Risley1974}
{Risley}, J.~S. \& {Geballe}, R. 1974, Phys. Rev. A, 9, 2485

\bibitem[{{Rose} {et~al.}(1958){Rose}, {Connor}, \& {Bastide}}]{Rose1958}
{Rose}, P.~H., {Connor}, R.~J., \& {Bastide}, R.~P. 1958, Bull. Am. Phys. Soc.,
  11, 3

\bibitem[{{Rudd} {et~al.}(1985{\natexlab{a}}){Rudd}, {Goffe}, {DuBois}, \&
  {Toburen}}]{Rudd1985_Hp_H2O}
{Rudd}, M.~E., {Goffe}, T.~V., {DuBois}, R.~D., \& {Toburen}, L.~H.
  1985{\natexlab{a}}, Phys. Rev. A, 31, 492

\bibitem[{{Rudd} {et~al.}(1985{\natexlab{b}}){Rudd}, {Goffe}, \&
  {Itoh}}]{Rudd1985_Hepp}
{Rudd}, M.~E., {Goffe}, T.~V., \& {Itoh}, A. 1985{\natexlab{b}}, Phys. Rev. A,
  32, 2128

\bibitem[{{Rudd} {et~al.}(1985{\natexlab{c}}){Rudd}, {Goffe}, {Itoh}, \&
  {Dubois}}]{Rudd1985_Hep_gases}
{Rudd}, M.~E., {Goffe}, T.~V., {Itoh}, A., \& {Dubois}, R.~D.
  1985{\natexlab{c}}, Phys. Rev. A, 32, 829

\bibitem[{{Rudd} {et~al.}(1985{\natexlab{d}}){Rudd}, {Itoh}, \&
  {Goffe}}]{Rudd1985_Hep}
{Rudd}, M.~E., {Itoh}, A., \& {Goffe}, T.~V. 1985{\natexlab{d}}, Phys. Rev. A,
  32, 2499

\bibitem[{{Rudd} {et~al.}(1985{\natexlab{e}}){Rudd}, {Kim}, {Madison}, \&
  {Gallagher}}]{Rudd1985_Hp_gases}
{Rudd}, M.~E., {Kim}, Y.-K., {Madison}, D.~H., \& {Gallagher}, J.~W.
  1985{\natexlab{e}}, Rev. Mod. Phys., 57, 965

\bibitem[{{Salazar-Zepeda} {et~al.}(2010){Salazar-Zepeda}, {Gleason},
  {Gonz\'{a}lez}, {Gonz\'{a}lez-Maga\~{n}a}, \& {Hinojosa}}]{Salazar2009}
{Salazar-Zepeda}, M.-H., {Gleason}, C., {Gonz\'{a}lez}, E.,
  {Gonz\'{a}lez-Maga\~{n}a}, O., \& {Hinojosa}, G. 2010, Nucl. Inst. Meth.
  Phys. Res. B, 268, 1558

\bibitem[{{Sataka} {et~al.}(1990){Sataka}, {Yagishita}, \&
  {Nakai}}]{Sataka1990}
{Sataka}, M., {Yagishita}, A., \& {Nakai}. 1990, J. Phys. B: At. Mol. Opt.
  Phys., 23, 1225

\bibitem[{{Schryber}(1967)}]{Schryber1967}
{Schryber}, U. 1967, Helvetica Physica Acta, 40, 1023

\bibitem[{{Schwadron} \& {Cravens}(2000)}]{Schwadron2000}
{Schwadron}, N.~A. \& {Cravens}, T.~E. 2000, \apj, 544, 558

\bibitem[{{Seredyuk} {et~al.}(2005){Seredyuk}, {McCullough}, {Tawara},
  {Gilbody}, {Bodewits}, {Hoekstra}, {Tielens}, {Sobocinski}, {Pesic},
  {Hellhammer}, {Sulik}, {Stolterfoht}, {Abu-Haija}, \&
  {Kamber}}]{Seredyuk2005}
{Seredyuk}, B., {McCullough}, R.~W., {Tawara}, H., {et~al.} 2005, Phys. Rev. A,
  71, 022705

\bibitem[{{Simon Wedlund} {et~al.}(2017){Simon Wedlund}, {Alho}, {Gronoff},
  {Kallio}, {Gunell}, {Nilsson}, {Lindkvist}, {Behar}, {Stenberg Wieser}, \&
  {Miloch}}]{CSW2017}
{Simon Wedlund}, C., {Alho}, M., {Gronoff}, G., {et~al.} 2017, A\&A, 604, A73

\bibitem[{{Simon Wedlund} {et~al.}(2019){Simon Wedlund}, {Behar}, {Kallio},
  {Nilsson}, {Alho}, {Gunell}, {Bodewits}, {Beth}, {Gronoff}, \&
  {Hoekstra}}]{CSW2018b}
{Simon Wedlund}, C., {Behar}, E., {Kallio}, E., {et~al.} 2019, accepted in
  A\&A, 1

\bibitem[{{Simon Wedlund} {et~al.}(2018){Simon Wedlund}, {Behar}, {Nilsson},
  {Alho}, {Kallio}, {Gunell}, {Bodewits}, {Heritier}, {Galand}, {Beth},
  {Rubin}, {Altwegg}, {Gronoff}, \& {Hoekstra}}]{CSW2018c}
{Simon Wedlund}, C., {Behar}, E., {Nilsson}, H., {et~al.} 2018, accepted in
  A\&A, 1

\bibitem[{{Simon Wedlund} {et~al.}(2016){Simon Wedlund}, {Kallio}, {Alho},
  {Nilsson}, {Stenberg Wieser}, {Gunell}, {Behar}, {Pusa}, \&
  {Gronoff}}]{CSW2016}
{Simon Wedlund}, C., {Kallio}, E., {Alho}, M., {et~al.} 2016, A\&A, 587, A154

\bibitem[{{Slavin} \& {Holzer}(1981)}]{Slavin1981}
{Slavin}, J.~A. \& {Holzer}, R.~E. 1981, J. Geophys. Res., 86, 11401

\bibitem[{{Thwaites}(1983)}]{Thwaites1983}
{Thwaites}, D.~I. 1983, Radiation Research, 95, 495

\bibitem[{{Toburen} \& {Nakai}(1969)}]{Toburen1969}
{Toburen}, L.~H. \& {Nakai}, M.~Y. 1969, Phys. Rev., 177, 191

\bibitem[{{Toburen} {et~al.}(1968){Toburen}, {Nakai}, \&
  {Langley}}]{Toburen1968}
{Toburen}, L.~H., {Nakai}, M.~Y., \& {Langley}, R.~A. 1968, Phys. Rev., 171,
  114

\bibitem[{{Toburen} \& {Wilson}(1977)}]{Toburen1977}
{Toburen}, L.~H. \& {Wilson}, W.~E. 1977, \jcp, 66, 5202

\bibitem[{{Toburen} {et~al.}(1980){Toburen}, {Wilson}, \&
  {Popowich}}]{Toburen1980}
{Toburen}, L.~H., {Wilson}, W.~E., \& {Popowich}, R.~J. 1980, Rad. Res., 82, 27

\bibitem[{{Tolstikhina} {et~al.}(2018){Tolstikhina}, {Imai}, {Winckler}, \&
  {Shevelko}}]{Tolstikhina2018}
{Tolstikhina}, I., {Imai}, M., {Winckler}, N., \& {Shevelko}, V. 2018, Basic
  Atomic Interactions of Accelerated Heavy Ions in Matter: Atomic Interactions
  of Heavy Ions, Springer Series on Atomic, Optical, and Plasma Physics
  (Springer International Publishing)

\bibitem[{{Uehara} \& {Nikjoo}(2002)}]{Uehara2002}
{Uehara}, S. \& {Nikjoo}, H. 2002, J. Phys. Chem. B, 106, 11051

\bibitem[{{Uehara} {et~al.}(2000){Uehara}, {Toburen}, {Wilson}, {Goodhead}, \&
  {Nikjoo}}]{Uehara2000}
{Uehara}, S., {Toburen}, L.~H., {Wilson}, W.~E., {Goodhead}, D.~T., \&
  {Nikjoo}, H. 2000, Rad. Phys. Chem., 59, 1

\bibitem[{{Van Zyl} \& {Stephen}(2014)}]{VanZyl2014}
{Van Zyl}, B. \& {Stephen}, T.~M. 2014, J. Geophys. Res. (Space Physics), 119,
  6925

\bibitem[{{Vech} {et~al.}(2015){Vech}, {Szego}, {Opitz}, {Kajdic}, {Fraenz},
  {Kallio}, \& {Alho}}]{Vech2015}
{Vech}, D., {Szego}, K., {Opitz}, A., {et~al.} 2015, J. Geophys. Res., 120,
  4613

\bibitem[{{Wegmann} \& {Dennerl}(2005)}]{Wegmann2005}
{Wegmann}, R. \& {Dennerl}, K. 2005, \aap, 430, L33

\bibitem[{{Werner} {et~al.}(1995){Werner}, {Beckord}, {Becker}, \&
  {Lutz}}]{Werner1995}
{Werner}, U., {Beckord}, K., {Becker}, J., \& {Lutz}, H.~O. 1995, Phys. Rev.
  Lett., 74, 1962

\bibitem[{{Williams} {et~al.}(1984){Williams}, {Geddes}, \&
  {Gilbody}}]{Williams1984}
{Williams}, I.~D., {Geddes}, J., \& {Gilbody}, H.~B. 1984, J. Phys. B At. Mol.
  Phys., 17, 1547

\bibitem[{{Williams}(1966)}]{Williams1966DC}
{Williams}, J.~F. 1966, Phys. Rev., 150, 7

\bibitem[{{Wittkower} \& {Betz}(1971)}]{Wittkower1971}
{Wittkower}, A.~B. \& {Betz}, H.~D. 1971, J. Phys. B: At. Mol. Phys., 4, 1173

\end{thebibliography}

\end{document}